\documentclass[twocolumn,aps,prb,notitlepage]{revtex4-2}
\pdfoutput=1
\usepackage{bm}
\usepackage{graphicx}
\usepackage{amssymb}
\usepackage{amsmath}
\usepackage{amsfonts}
\usepackage{upgreek}
\usepackage{comment}
\usepackage[margin=.7in]{geometry}
\usepackage[euler]{textgreek}
\usepackage{lineno}
\usepackage{braket}
\usepackage[caption=false]{subfig}
\usepackage{floatrow}
\usepackage{gensymb}
\usepackage{multirow}
\usepackage{wasysym}
\usepackage{mathtools}
\usepackage[usenames,dvipsnames]{xcolor}
\usepackage[normalem]{ulem}
\usepackage[colorlinks]{hyperref}
\hypersetup{
    colorlinks=true,
    citecolor=blue,
    linkcolor=blue,
    filecolor=magenta,      
    urlcolor=blue,
    }
\newcommand{\stkout}[1]{\ifmmode\text{\sout{\ensuremath{#1}}}\else\sout{#1}\fi}
\newcommand{\bs}{\boldsymbol}
\newcommand{\pd}{\partial}
\newcommand{\dprime}{\prime\prime}
\newcommand{\tprime}{\prime\prime\prime}

\begin{document}

\title{Quantum unidirectional magnetoresistance}
\author{M. Mehraeen}
\email{mxm1289@case.edu}
\author{Pengtao Shen}
\author{Steven S.-L. Zhang}
\email{shulei.zhang@case.edu}
\affiliation{Department of Physics, Case Western Reserve University, Cleveland, OH 44106, USA}
\date{\today}
\begin{abstract}
We predict unidirectional magnetoresistance effects arising in a bilayer composed of a nonmagnetic metal and a ferromagnetic insulator, whereby both longitudinal and transverse resistances vary when the direction of the applied electric field is reversed or the magnetization of the ferromagnetic layer is rotated. In the presence of spin-orbit coupling, an electron wave incident on the interface of the bilayer undergoes a spin rotation and a momentum-dependent phase shift. Quantum interference between the incident and reflected waves furnishes the electron with an additional velocity that is even in the in-plane component of the electron's wavevector, giving rise to quadratic magnetotransport that is rooted in the wave nature of electrons. The corresponding unidirectional magnetoresistances exhibit decay lengths at the scale of the Fermi wavelength--distinctive signatures of the quantum nonlinear magnetotransport effect.
\end{abstract}

\maketitle

\section{Introduction}

As fingerprints of electron waves, quantum interference effects in electron transport have been of fundamental interest \cite{Anderson58pr_localization,AbrahamsPRL79_WL-2D,BERGMANN84Phys-rep_WL-thin-film,Webb85PRL_AB-oscillation_NM-ring,Bachtold99_AB-osc_Carbon-NT,Peng10NM_AB-TI,Spivak10RMP_QM-transport_2DEF}. And from the perspective of applications, studies of coherent quantum transport of electrons--and more generally information carriers--may also underpin the development
of future quantum devices, including quantum computers \cite{Whiticar20NC_Majorana-AB}. However, interference-based magnetoresistances have thus far been limited to the \textit{linear} response regime \cite{AbrahamsPRL79_WL-2D,BERGMANN84Phys-rep_WL-thin-film,altshuler1980prb,hikami1980}, where they remain \textit{invariant} under the reversal of the applied magnetic field.

Recently, there has been increasing interest in an emergent unidirectional
magnetoresistance (UMR) effect observed in various bilayer systems composed of a nonmagnetic layer and a ferromagnetic layer ~\cite{avci2015natphys,Olejnik15PRB_UMR-semicond,yasuda2016prl,lv2018unidirectional,guillet2021prb,liu2021magnonic,mehraeen2022spin}. This novel nonlinear magnetotransport effect features a variation in  magnetoresistance when the direction of the applied
electric field is reversed, at variance with common linear magnetoresistances, which are electric-field independent. 

While the list of systems that can host such UMRs
keeps growing, they consistently possess three essential ingredients: strong
spin-orbit coupling (SOC), structural inversion asymmetry, and broken time reversal symmetry. In
terms of microscopic mechanisms, lying at the heart of UMR effects are the
spin-momentum coupling and spin-asymmetry in electron scattering~\cite%
{shulei2016prb,Langenfeld16APL_UMR-FMR,yasuda2016prl,avci2018prl,guillet2021prb}. Quantum interference, however,
has not been demonstrated to play any role in nonlinear magnetotransport.

\begin{figure}[tph]
    \includegraphics[width=1.1\linewidth]{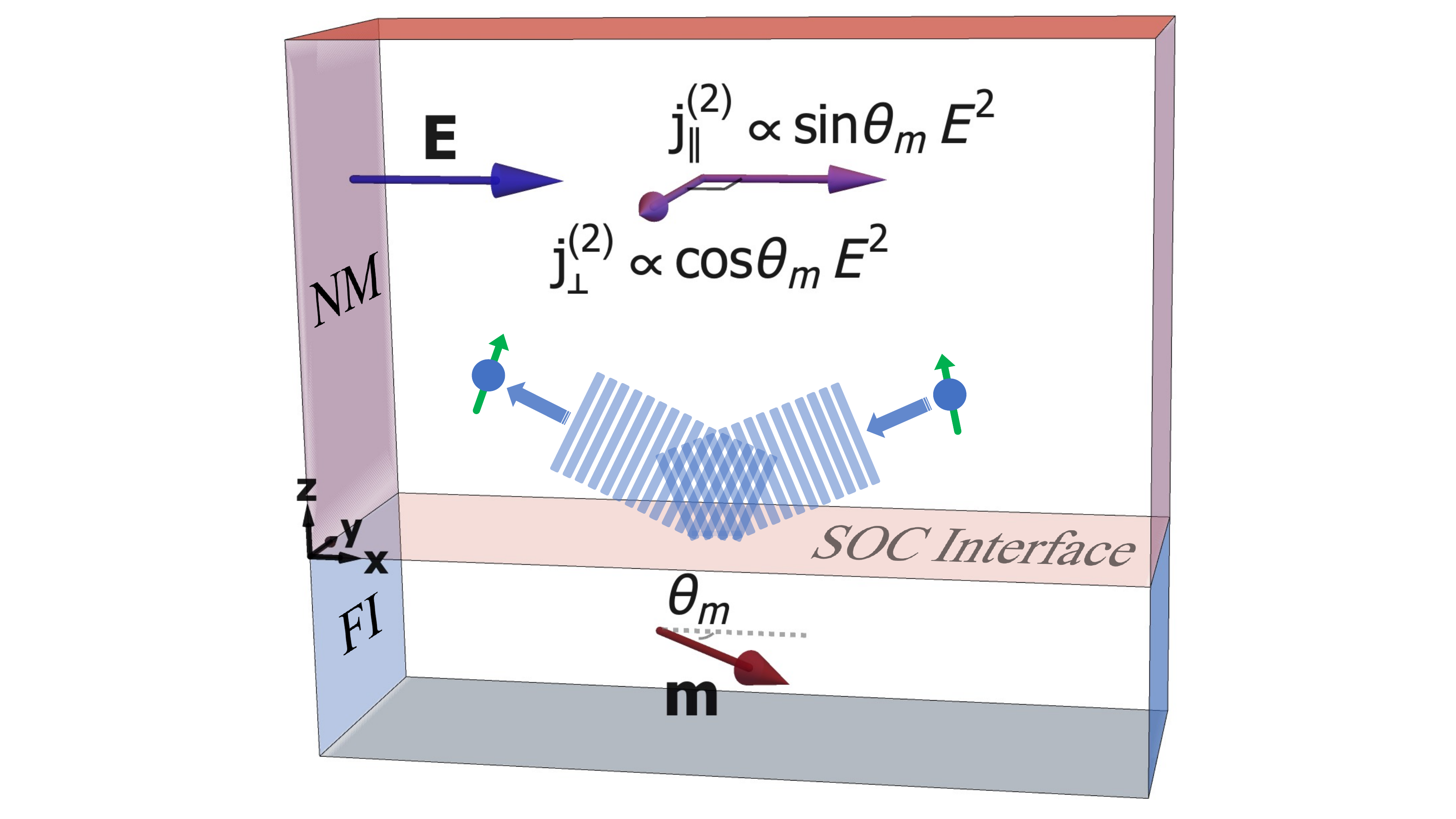}
    \caption{Schematic of the quantum unidirectional
magnetoresistance (QUMR) effect in a NM$|$FI bilayer. An electron in the NM layer scatters off the interface with spin-orbit coupling (SOC), resulting in an interference between the incident and reflected waves, which in turn gives rise to a nonlinear current. Both longitudinal and transverse components of the nonlinear current, $j_{\parallel}^{(2)}$ and $j_{\perp}^{(2)}$, are sensitive to the orientation of the in-plane magnetization $\mathbf{m}$ relative to the electric field $\mathbf{E}$, with different dependences on $\theta_m$--the angle between $\mathbf{m}$ and $\mathbf{E}$.}
    \label{fig1}%
\end{figure}

In this work, we unveil the role of quantum interference in generating both
\textit{longitudinal} and \textit{transverse} UMRs in bilayer structures
consisting of a nonmagnetic metal (NM) and a ferromagnetic insulator (FI), in the presence of
SOC at the interface. Microscopically, the scattering of an
electron wave at such an interface hinges on the wavevector and spin
orientations of the incident electron, as shown schematically in Fig.~\ref{fig1}.
Due to the interference between the incident and reflected waves, the
electron acquires an additional velocity that is \textit{even}
in the in-plane component of the wavevector (as we will show explicitly),
which in turn gives rise to a nonlinear current density of the following
form:
\begin{equation}
\mathbf{j}^{(2)}=\sigma _{\parallel }^{(2)}\mathbf{z}\cdot \left( \mathbf{E}%
\times \mathbf{m}\right) \;\mathbf{E}+\sigma _{\bot }^{(2)}\mathbf{E}\cdot
\mathbf{m}\;\mathbf{z}\times \mathbf{E},  \label{vectorform}
\end{equation}%
where $\mathbf{E}$ is the applied electric field, $\mathbf{m}$ and $\mathbf{%
z~}$are two unit vectors denoting, respectively, the directions of
magnetization of the ferromagnetic layer and the\textbf{\ }normal to the
bilayer interface. $\sigma _{\parallel }^{(2)}$ and $\sigma _{\perp }^{(2)}$, both being $\mathbf{E}$-independent, characterize, respectively, the strengths of
the longitudinal and transverse contributions to the nonlinear transport, with
the superscripts indicating that $\mathbf{j}^{(2)}$ is of the second order
in the electric field. 

Given the quantum mechanical origin of the phenomena, we shall refer to the corresponding UMRs as \textit{quantum unidirectional
magnetoresistances} (QUMRs), in order to distinguish them from their semiclassical counterparts that have heretofore been reported.

\section{The Model}

For simplicity, we shall restrict ourselves to the
nonlinear magnetotransport in a NM$|$FI bilayer, whereby electrons incident on the interface are completely reflected back into the
NM layer and hence the charge transport in the ferromagnetic layer can be disregarded. Furthermore, we assume the presence of an exchange interaction coupling the electron spin and the interfacial magnetization of the FI and a Rashba-type SOC arising from the structural inversion asymmetry at the interface. Consider a setup described by Fig.~\ref{fig1}, with a three-dimensional (3D) electron gas contained in a semi-infinite NM layer occupying the $z>0$ half-space and a FI layer occupying the other half.

The 3D electron gas can be described by the Hamiltonian \cite{Stiles13PRB_SOT,tokatly2015prb}
\begin{equation}
\label{Ham}
\hat{H}=\frac{\hat{\mathbf{p}}^{2}}{2m}
+
\frac{\alpha _{R}}{\hbar
}\delta (z)\hat{\boldsymbol{\sigma }}\cdot (\hat{\mathbf{p}}\times \mathbf{z}%
)-J_{ex}\delta (z)\hat{\boldsymbol{\sigma }}\cdot \mathbf{m}+V_{b}\Theta
(-z), 
\end{equation}%
where $\alpha _{R}$ and $J_{ex}$ are the coefficients of the interfacial Rashba SOC and exchange interaction, respectively, and $V_{b}$ is the height of the energy barrier,
which is greater than the Fermi energy of electrons in the NM layer.

The general scattering state may be written as
\begin{equation}
\label{gen_scat}
\boldsymbol{\psi }_{\text{scat.}}=\left\{
\begin{array}{c}
e^{i\mathbf{q}\cdot \boldsymbol{\rho }}\left( e^{-ik_{z}z}+e^{ik_{z}z}\hat{R}_{\mathbf{q}}%
\right) \boldsymbol{\chi },\;\;z>0\;\; \\
e^{i\mathbf{q}\cdot \boldsymbol{\rho }}e^{\kappa_z z}\;\hat{T}_{\mathbf{q}}\boldsymbol{\chi
},\;\;z<0\;\;%
\end{array},\right.
\end{equation}%
where $\mathbf{q}\left[ =\left( k_{x},k_{y}\right) \right] $ and $%
\boldsymbol{\rho }\left[ =\left( x,y\right) \right] $ are the wave- and
position-vector in the $x$-$y$ plane, wherein the system is translationally
invariant and allows the propagation of plane waves, $k_{z}$ is the $z$%
-component of the wavevector of the propagating wave in the NM layer, and $%
\kappa_z^{-1}\left[ =\left( 2m V_{b}/\hbar ^{2}-k_{z}^{2}\right)
^{-1/2}\right] $ characterizes the decay length of the evanescent wave in
the FI layer. $\hat{R}_{\mathbf{q}}$ and $\hat{T}_{\mathbf{q}}$ are $2\times 2$ matrices in spin
space, which describe, respectively, the spin-dependent reflection and
transmission amplitudes, and the spinors $\boldsymbol{\chi }$ are taken to
be the eigenstates satisfying $\hat{\boldsymbol{\sigma }}\cdot \mathbf{m}\;\chi ^{\pm }=\pm
\chi ^{\pm }$.

By imposing the standard boundary conditions at the Rashba interface, namely the continuity of the wavefunction and the discontinuity of its spatial derivative along the $z$ direction--as detailed in Appendix~\ref{appendixA}--we find 
\begin{equation}
\label{eq:R_q}
\hat{R}_{\mathbf{q}}
=
e^{i \left( \varphi_{\mathbf{q}} + \vartheta _{\mathbf{q}}
\hat{\boldsymbol{\sigma}}\cdot \mathbf{n}_{\mathbf{q}} \right)},
\end{equation}
and $\hat{T}_{\mathbf{q}}=1+\hat{R}_{\mathbf{q}}$, where $\varphi _{\mathbf{q}}=\arcsin \left( 2k_{z}\kappa_z /\varkappa_{\mathbf{q}}\right)$, $\vartheta _{\mathbf{q}}=\arcsin \left( 2k_{z}Q_{\mathbf{q}}/\varkappa_{\mathbf{q}}\right) $, $\varkappa_{\mathbf{q}}^{2}=\left[ Q_{%
\mathbf{q}}^{2}-(\kappa_z^{2}+k_{z}^{2})\right] ^{2}+\left( 2k_{z}Q_{\mathbf{q%
}}\right) ^{2}$, and $\mathbf{n}_{\mathbf{q}} \mathbf{=\mathbf{Q}_{%
\mathbf{q}}/}Q_{\mathbf{q}}$. Here, $\mathbf{Q}_{\mathbf{q}}\equiv \eta _{R}%
\mathbf{q}\times \mathbf{z}-\xi _{ex}\mathbf{m}$, with $\eta _{R}\equiv
2m \alpha _{R}/\hbar ^{2}$ a dimensionless constant characterizing the strength of the interfacial Rashba SOC and $\xi _{ex}\equiv 2m J_{ex}/\hbar ^{2}$ the rescaled exchange interaction. Note
that $\hat{R}_{\mathbf{q}}$ is unitary, as enforced by the conservation of probability
flux.

All key information about the reflected wave, such
as the phase shift, in particular, is encapsulated in Eq.~(\ref{eq:R_q}), from which the physical meaning of $\hat{R}_{\mathbf{q}}$ is evident: The angle $\varphi_{\mathbf{q}}$ describes a largely spin-independent phase shift, in conjunction with a spin rotation by an angle $\vartheta_{\mathbf{q}}$ about a rotation axis taken in the direction of $\mathbf{n}_{\mathbf{q}}$. Note that both $\vartheta _{\mathbf{q}}$ and $\mathbf{n}_{\mathbf{q}}$ depend, through the vector field $\mathbf{Q}_{\mathbf{q}}$, on the wavevector $\mathbf{q}$ and the magnetization $\mathbf{m}$, as the spin rotation is brought about by the interfacial exchange interaction and Rashba SOC.

\begin{figure}[tph]
\sidesubfloat[]{\includegraphics[width=0.75\linewidth,trim={1.5cm 2cm 1cm 1cm}]{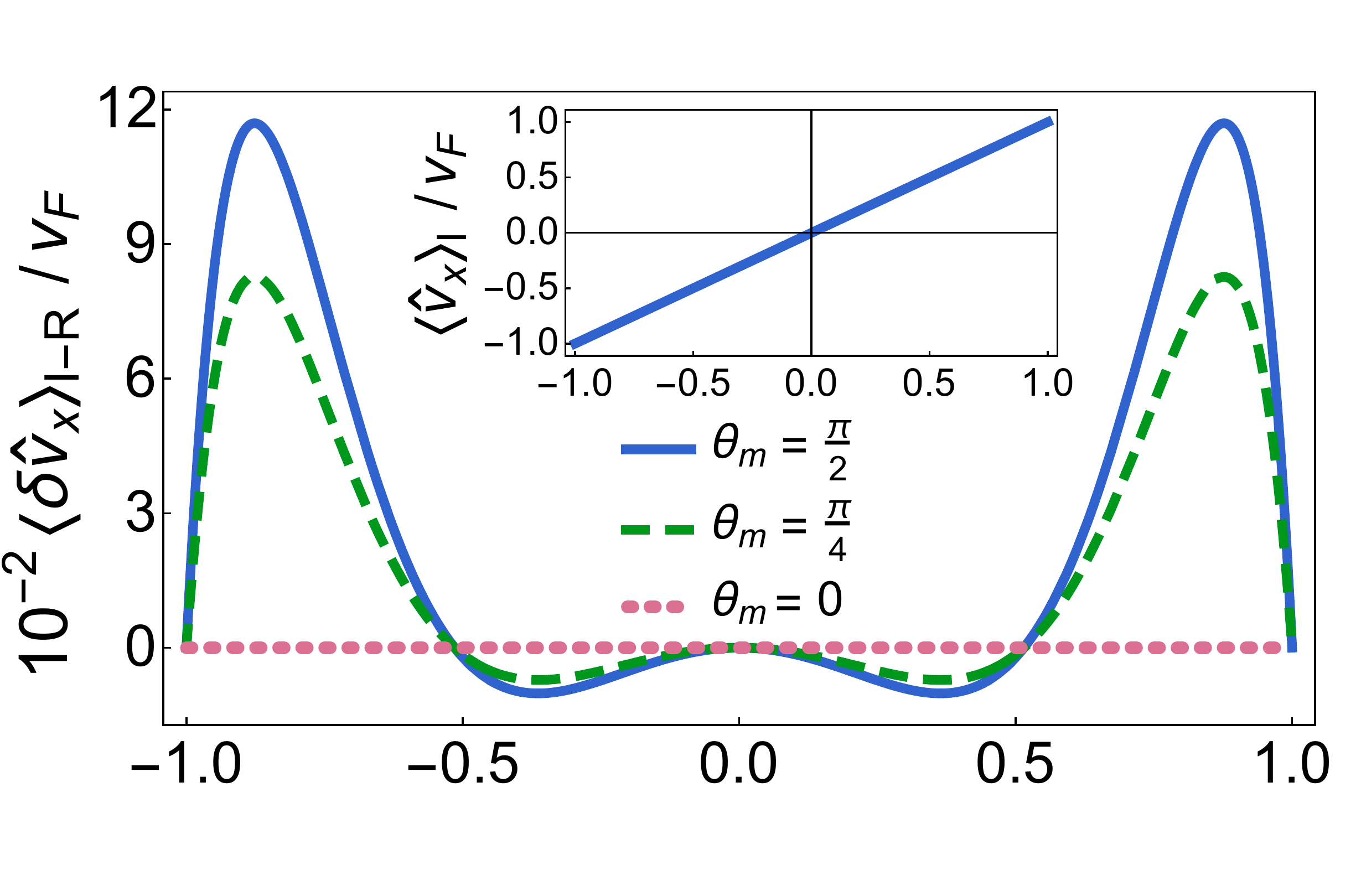}\label{fig:S2_a}}\quad%
    \\
\sidesubfloat[]{\includegraphics[width=0.75\linewidth,trim={2.5cm 0.5cm 0.5cm 1.5cm}]{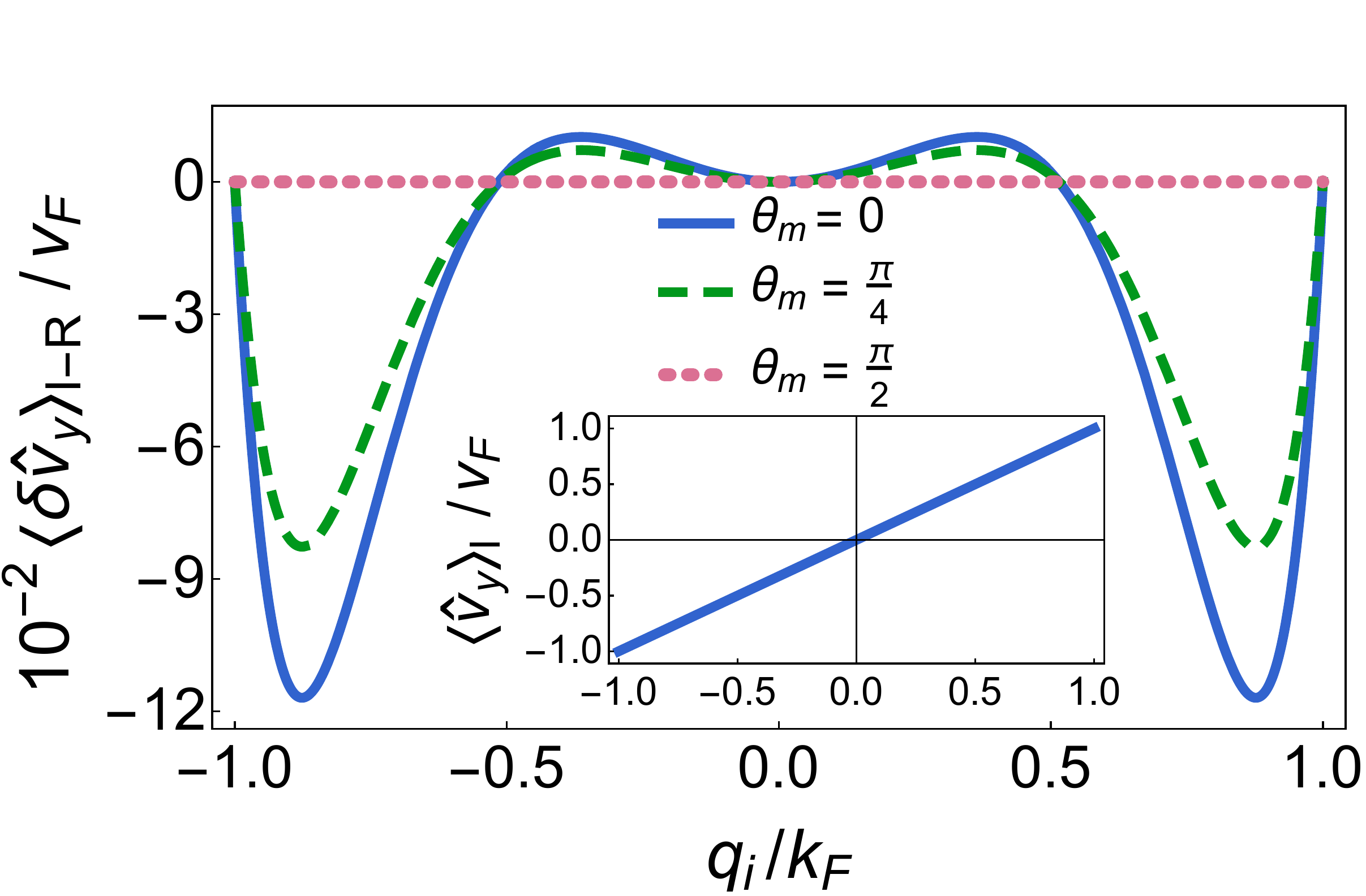}\label{fig:S2_b}}%
    \caption{The even part of the spatial average of $\langle \hat{\boldsymbol{v}}_{\parallel}\left( \mathbf{q}%
\right) \rangle _{I\text{-}R}$, given by $\langle \protect\delta \hat{%
\boldsymbol{v}}_{\parallel}\rangle _{I\text{-}R}=\frac{1}{2}\left[ \langle \hat{\boldsymbol{%
v}}_{\parallel}(\mathbf{q})\rangle _{I\text{-}R}+\langle \hat{\boldsymbol{v}}_{\parallel}(-\mathbf{q}%
)\rangle _{I\text{-}R}\right] $, as a function of in-plane momentum for three different values of $\theta_m$. (a) Plots
of $\langle \protect\delta \hat{v}_{x}\rangle _{I\text{-}R}$ (main) and $%
\langle \hat{v}_{x}\rangle _{I}$ (inset) as functions of $q_i = q_{x}$ at $q_{y}=0$. (b) Plots of $\langle \protect\delta \hat{v}_{y}\rangle _{I\text{-}R}$
(main) and $\langle \hat{v}_{y}\rangle _{I}$ (inset) as functions of $q_i=q_{y}$ at $q_{x}=0$.}
    \label{fig2}
\end{figure}

\section{Interference velocity}

To provide a heuristic picture of the QUMR effect, let us first consider the spin-averaged velocity relevant to the charge transport in the NM layer, which is given by the expectation value of the velocity operator with
respect to the spinor part of the scattering state for $z>0$, \textit{i.e.},
\begin{equation}  
\label{Eq:v}
\langle \hat{\boldsymbol{v}}\rangle \equiv \sum_{\sigma }\text{Re}\left[ \left(
\psi _{\text{scat.}}^{\sigma }\right) ^{\dag }\frac{\hat{\mathbf{p}}}{%
m}\left( \psi _{\text{scat.}}^{\sigma }\right) \right] =\langle
\hat{\boldsymbol{v}}\rangle _{I}+\langle \hat{\boldsymbol{v}}\rangle
_{R}+\langle \hat{\boldsymbol{v}}\rangle _{I\text{-}R},
\end{equation}%
where the sum runs over the spin index. This velocity can be decomposed into
three parts, as shown in Eq.~(\ref{Eq:v}). The first two terms, $\langle \hat{%
\boldsymbol{v}}\rangle _{I}$ and $\langle \hat{\boldsymbol{v}}\rangle _{R}$,
are associated with the incident and reflected waves, respectively, and more
intriguingly, there exists another term, $\langle \hat{\boldsymbol{v}}%
\rangle _{I\text{-}R}$, originating from the interference between the
incident and the reflected waves. The in-plane component of this \textit{%
interference velocity} can be expressed as
\begin{equation}  \label{Eq:v_I-R}
\langle \hat{\boldsymbol{v}}_{\parallel }\rangle _{I\text{-}R}=\frac{4\hbar
\mathbf{q}}{m} \cos \left(\vartheta _{\mathbf{q}}%
\right) \cos \left( 2k_{z}z+\varphi _{\mathbf{q}}\right),
\end{equation}
where $\hat{\boldsymbol{v}}_{\parallel }=\left( \hat{v}_{x},\hat{v}%
_{y}\right) $.

There are two crucial differences between the interference velocity $\langle
\hat{\boldsymbol{v}}\rangle _{I\text{-}R}$ and the other counterparts, as
displayed in Fig.~\ref{fig2}. While $\langle \hat{\boldsymbol{v}}%
\rangle _{I}$ and $\langle \hat{\boldsymbol{v}}\rangle _{R}$ share the same
in-plane component $\frac{\hbar \mathbf{q}}{m}$ (which is
clearly odd in $\mathbf{q}$), the interference velocity $\langle \hat{%
\boldsymbol{v}}\rangle _{I\text{-}R}$ includes a constituent that is \textit{%
even} in $\mathbf{q}$~\footnote{The interference velocity also has a component that is odd in $\mathbf{q}$, which contributes to an anisotropic magnetoresistance--in linear response to the applied electric field--with an angular dependence similar to the spin-Hall magnetoresistance~\cite{Saitoh13PRL_SH-MR,xJiang14prb_Rashba-AMR,slzhang15PRB_AMR}}--which we define as $\langle \delta \hat{\boldsymbol{%
v}}_{\parallel }\rangle _{I\text{-}R}\equiv \frac{1}{2}\left[ \langle \hat{%
\boldsymbol{v}}_{\parallel }\left( \mathbf{q}\right) \rangle _{I\text{-}%
R}+\langle \hat{\boldsymbol{v}}_{\parallel }\left( -\mathbf{q}\right)
\rangle _{I\text{-}R}\right] $--due to the combined action of interference
and spin-dependent scattering at the interface, as long as the magnetization
$\mathbf{m}$ is \textit{not }perpendicular to the $x$-$y$ plane.

Furthermore, $\langle \hat{\boldsymbol{v}}\rangle _{I}$ and $\langle \hat{%
\boldsymbol{v}}\rangle _{R}$ are inert to the rotation of the magnetization $%
\mathbf{m}$, but $\langle \hat{\boldsymbol{v}}_{\parallel }\rangle _{I\text{-%
}R}$ is \textit{not}. In fact, the latter is exquisitely sensitive to the
variation in the angle formed by $\mathbf{m}$ and $\mathbf{q}$, with $%
\langle \delta \hat{v}_{x}\rangle _{I\text{-}R}$ and $\langle \delta \hat{v}%
_{y}\rangle _{I\text{-}R}$ exhibiting different angular dependences. It is
these distinctive features of the interference velocity that give rise to the QUMR, as we will evaluate below.

\section{Disorder scattering}

To take into account interfacial disorder--which leads to diffusive scattering of incident electrons, we add a random impurity potential that is localized at the interface, i.e.,  $\hat{H} \rightarrow \hat{H} + V^{\text{imp}}(\mathbf{r})$ with $V^{\text{imp}}(\mathbf{r})=V^{\text{imp}}(\bs{\rho})\delta(z)$, leading to the Dyson equation
\begin{equation}
\label{G_tilde}
\hat{\tilde{G}} \left(\mathbf{r},\mathbf{r}^{\prime}\right)
=
\hat{G}\left(\mathbf{r},\mathbf{r}^{\prime}\right)
+
\int d\mathbf{r}_1
\hat{G} \left(\mathbf{r},\mathbf{r}_1\right) 
V^{\text{imp}}\left({\mathbf{r}}_1\right)
\hat{\tilde{G}} \left(\mathbf{r}_1,\mathbf{r}^{\prime}\right),
\end{equation}
for the disordered propagator $\hat{\tilde{G}}$. And we assume the impurity potential has the white noise distribution $\braket{V^{\text{imp}} \left(\bs{\rho}\right)}
= 0$ and $\braket{V^{\text{imp}} \left(\bs{\rho}\right) V^{\text{imp}} \left(\bs{\rho}^{\prime}\right)} = (\hbar^2/2m)^2 \eta_{\gamma} \delta\left(\bs{\rho} - \bs{\rho}^{\prime}\right)$, where the dimensionless parameter $\eta_{\gamma}$ characterizes the strength of the impurity interaction and $\braket{\cdots}$ here denotes the configurational average over impurity positions.

Taking the disorder average of Eq.~(\ref{G_tilde}) for the retarded propagator and introducing the decomposition 
$\braket{\hat{\tilde{G}}^R (\mathbf{r},\mathbf{r}^{\prime};\epsilon)}
=
\int_{\mathbf{q}} e^{i \mathbf{q} \cdot (\bs{\rho} - \bs{\rho}^{\prime} )} \hat{\tilde{g}}^R_{\mathbf{q}} (z,z^{\prime};\epsilon)$ with $\int_{\mathbf{q}} \equiv \int \frac{d^2\mathbf{q}}{\left( 2\pi \right)^2}$, the solution of the dressed propagator at the interface ($z,z^{\prime}=0$) reads $\hat{\tilde{g}}^R
=
[(\hat{g}^R)^{-1} - \hat{\tilde{\Sigma}}^{\text{imp}}]^{-1}$, where $\hat{g}^R$ is the bare propagator corresponding to Eq.~(\ref{Ham}) and $\hat{\tilde{\Sigma}}^{\text{imp}} \left(\epsilon\right) = (\hbar^2/2m)^2 \eta_{\gamma} \int_{\mathbf{q}}
\hat{g}^R_{\mathbf{q}} \left(0,0;\epsilon\right)$ is the configurationally averaged interfacial self-energy. Once $\hat{\tilde{g}}^R$ is known at the interface, the dressed reflection matrix is obtained, which, using the decomposition $\hat{\tilde{\Sigma}}^{\text{imp}} \left(\epsilon\right) = \hbar^2 \tilde{\xi}_{\mu}\left(\epsilon\right) \hat{\sigma}^{\mu} /2m$ with $\mu = 0,x,y,z$, and following the steps provided in Appendix~\ref{appendixA}, may be expressed as
\begin{equation}
\label{R_tilde}
\hat{\tilde{R}}_{\mathbf{q}}
=
\frac{e^{i \tilde{\varphi}_{\mathbf{q}}}}{\tilde{\varkappa}_{\mathbf{q}}}
\left(
\nu_{\mathbf{q}}
e^{i \tilde{\vartheta} _{\mathbf{q}}
\hat{\boldsymbol{\sigma}}\cdot
\tilde{\mathbf{n}}_{\mathbf{q},\text{R}}}
+
i \nu^{\prime}_{\mathbf{q}}
e^{i \tilde{\vartheta}^{\prime} _{\mathbf{q}}
\hat{\boldsymbol{\sigma}}\cdot 
\tilde{\mathbf{n}}_{\mathbf{q},\text{I}}}
\right),
\end{equation}
where 
$\tilde{\varphi}_{\mathbf{q}}
=
\arcsin[2 (\tilde{\kappa}_z \tilde{k}_z 
+
\tilde{\mathbf{Q}}_{\mathbf{q},\text{R}}
\cdot
\tilde{\mathbf{Q}}_{\mathbf{q},\text{I}}) / \tilde{\varkappa}_{\mathbf{q}}]$, 
$\tilde{\vartheta}_{\mathbf{q}}
=
\arcsin(2 k_z \tilde{Q}_{\mathbf{q},\text{R}}/\nu_{\mathbf{q}})$ and
$\tilde{\vartheta}_{\mathbf{q}}^{\prime} 
= 
\arcsin(2 k_z \tilde{Q}_{\mathbf{q},\text{I}}/\nu_{\mathbf{q}}^{\prime})$ with
$\tilde{\varkappa}_{\mathbf{q}}^2
=
(\tilde{Q}_{\mathbf{q},\text{R}}^2 
-
\tilde{Q}_{\mathbf{q},\text{I}}^2
-
\tilde{\kappa}_z^2 + \tilde{k}_z^2)^2
+
4(\tilde{\mathbf{Q}}_{\mathbf{q},\text{R}} 
\cdot
\tilde{\mathbf{Q}}_{\mathbf{q},\text{I}}
+
\tilde{\kappa}_z \tilde{k}_z)^2$, 
$\nu_{\mathbf{q}}^2 
=
[\tilde{Q}_{\mathbf{q},\text{R}}^2 
-
\tilde{Q}_{\mathbf{q},\text{I}}^2
-
\tilde{\kappa}_z^2 - \tilde{k}_z^2
-
2 \tilde{k}_z \text{Im}(\tilde{\xi}_{0})]^2
+
(2 k_z \tilde{Q}_{\mathbf{q},\text{R}})^2$ and 
$\nu_{\mathbf{q}}^{\prime 2} 
=
4[ \tilde{\mathbf{Q}}_{\mathbf{q},\text{R}} 
\cdot
\tilde{\mathbf{Q}}_{\mathbf{q},\text{I}}
-
\tilde{\kappa}_z \text{Im}(\tilde{\xi}_{0}
)]^2
+
(2 k_z \tilde{Q}_{\mathbf{q},\text{I}})^2$. Here, we have introduced the disordered quantities 
$\tilde{k}_z \equiv k_z - \text{Im}(\tilde{\xi}_0)$, 
$\tilde{\kappa}_z \equiv \kappa_z +\text{Re}(\tilde{\xi}_0)$ and $\tilde{\mathbf{Q}}_{\mathbf{q}} \equiv \mathbf{Q}_{\mathbf{q}} + \tilde{\bs{\xi}}$ with 
$\tilde{\mathbf{Q}}_{\mathbf{q}, \text{R}}
=
\text{Re}(\tilde{\mathbf{Q}}_{\mathbf{q}})$, 
$\tilde{\mathbf{Q}}_{\mathbf{q}, \text{I}}
=
\text{Im}(\tilde{\mathbf{Q}}_{\mathbf{q}})$, 
$\tilde{\mathbf{n}}_{\mathbf{q},\text{R}}
=
\tilde{\mathbf{Q}}_{\mathbf{q},\text{R}}/
\tilde{Q}_{\mathbf{q},\text{R}}$ and 
$\tilde{\mathbf{n}}_{\mathbf{q},\text{I}}
=
\tilde{\mathbf{Q}}_{\mathbf{q},\text{I}}/
\tilde{Q}_{\mathbf{q},\text{I}}$. Note that in the limit of vanishing disorder, $\nu_{\mathbf{q}}, \tilde{\varkappa}_{\mathbf{q}} \rightarrow \varkappa_{\mathbf{q}}$, while $\nu_{\mathbf{q}}^{\prime} \rightarrow 0$ (hence $\hat{\tilde{R}}_{\mathbf{q}} \rightarrow \hat{R}_{\mathbf{q}}$). We thus see that the overall effect of the interfacial disorder is a modulation of both the amplitude and phase of the reflection matrix.

Along with the surface disorder, bulk impurities are also present, whose contribution to the propagator may be included through an additional local self-energy $\Sigma^B(\mathbf{r})$. This results in a local scattering time  $\tau(\mathbf{r})= - \hbar/2 \text{Im} [\Sigma^B(\mathbf{r})]$ and a reduction of the propagator as \cite{camblong1994theory}
\begin{equation}
\hat{\tilde{G}} \left(\mathbf{r},\mathbf{r}^{\prime}\right)
\rightarrow
\hat{\tilde{G}} \left(\mathbf{r},\mathbf{r}^{\prime}\right) \text{exp} \left[- \int \limits_{\Gamma\left[\mathbf{r}, \mathbf{r}^{\prime}\right]} \frac{ds^{\dprime}}{2 l\left(\mathbf{r}^{\dprime}\right)} \right],
\end{equation}
which arises as a result of the damping of the electron wavefunction as it propagates along the straight path $\Gamma\left[\mathbf{r}, \mathbf{r}^{\prime}\right]$ connecting the points $\mathbf{r}$ and $\mathbf{r}^{\prime}$, and accounts for the average scattering encountered by the electron. Here, $l(\mathbf{r})=v_F \tau(\mathbf{r})$ is the local scattering length of the conduction electrons. In the NM layer, the loss of momentum associated with the damping of the wavefunction corresponds to the replacement $k_z \rightarrow k_z \sqrt{1 + i k_F/k_z^2 l_n}$ at the Fermi level, where $l_n$ is the mean free path.

\section{QUMR}

\begin{figure}[tph]
    \sidesubfloat[]{\includegraphics[width=0.88\linewidth,trim={1.5cm 0.7cm 0.5cm 0}]{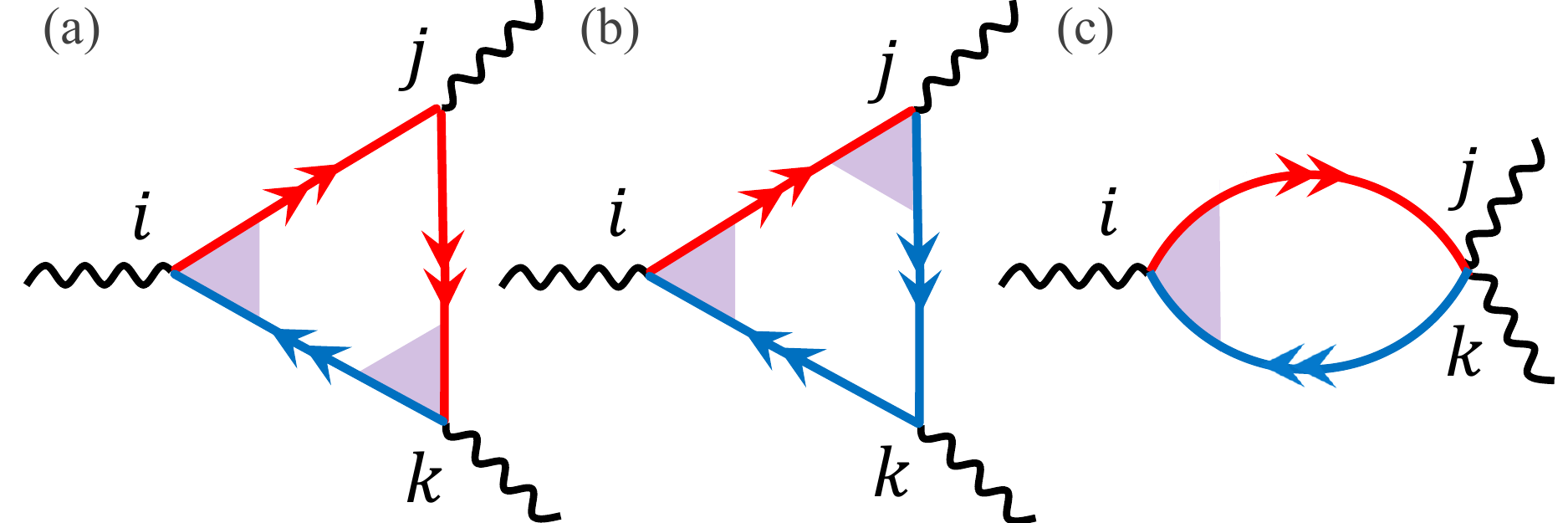}}
	\caption{Diagrammatic structure of the quadratic response. Together, (a-c) and their $j \leftrightarrow k$ counterparts comprise the dressed conductivity $\tilde{\sigma}_{ijk}$. The red (blue) double-arrowed lines represent dressed retarded (advanced) propagators, while the purple-shaded areas are dressed vertices.}
    \label{fig3}
\end{figure}

To capture the nonlinear transport, we evaluate the relevant quadratic Kubo formulas. The nonlinear conductivity is then composed of two parts, $\tilde{\sigma}_{ijk}= \tilde{\sigma}_{ijk}^{(a)} + \tilde{\sigma}_{ijk}^{(b)}$, which diagrammatically correspond to dressed triangle and three-photon bubble diagrams \cite{parker2019diagrammatic, du2021quantum, rostami2021gauge}, as shown in Fig.~\ref{fig3}. In general, the nonlinear conductivities will be modified by both self-energy and vertex corrections. However, as explained in Appendix~\ref{appendixB}, the contributions of the latter to the total conductivities turn out to be negligible in the weak disorder limit. Thus, to the leading order in the interfacial disorder, the conductivities may be expressed as
\begin{subequations}
\label{eq:sigma_ijk}
\begin{align}
\tilde{\sigma}_{ijk}^{(a)}
&=
\frac{2 e^3 \hbar^5}{\pi m^{3}} \int \limits_{S^I} q_i q_j q_k
\text{Im} \left[ \text{Tr} \left(
\pd_{\epsilon_F} \hat{\tilde{g}}^{R}_{\mathbf{q}, z z^{\prime}} 
\hat{\tilde{g}}^{R}_{\mathbf{q}, z^{\prime} z^{\dprime}} 
\hat{\tilde{g}}^{A}_{\mathbf{q}, z^{\dprime} z}
\right) \right],
\\
\tilde{\sigma}_{ijk}^{(b)}
&=
\frac{e^3 \hbar^3}{\pi m^2} \int \limits_{S^{II}} q_i \delta_{jk}
\text{Im} \left[ \text{Tr} \left(
\pd_{\epsilon_F} \hat{\tilde{g}}^{R}_{\mathbf{q}, z z^{\prime}} 
\hat{\tilde{g}}^A_{\mathbf{q}, z^{\prime} z}
\right) \right],
\end{align}
\end{subequations}
where $\int_{S^I} \equiv \int_{\mathbf{q}} 
\int_0^{\infty} dz^{\prime} \int_0^{\infty} dz^{\dprime}$, $\int_{S^{II}} \equiv \int_{\mathbf{q}} 
\int_0^{\infty} dz^{\prime}$, $\hat{\tilde{g}}^R_{\mathbf{q}, z z^{\prime}} \equiv \hat{\tilde{g}}^R_{\mathbf{q}} (z,z^{\prime};\epsilon_F)$ and $\hat{\tilde{g}}^A_{\mathbf{q}, z z^{\prime}} = (\hat{\tilde{g}}^R_{\mathbf{q}, z^{\prime} z})^{\dagger}$ is the advanced propagator.

Without loss of generality, let us set the electric field $\mathbf{E}$ in the $\mathbf{x}$ direction, so that we are only concerned with two tensor elements:  $\sigma _{xxx}$ and $\sigma _{yxx}$, governing the
nonlinear charge transport in the longitudinal and transverse directions, respectively. One can show that $\sigma _{xxx}$ must vanish when $\mathbf{m}$ is parallel to the $x$ axis, whereas $\sigma _{yxx}$ has to be zero when $\mathbf{m}$ is parallel to the $y$ axis (which is perpendicular to $\mathbf{E}$). These results can be understood intuitively by the following symmetry analysis. When $\mathbf{m}$ is parallel to the $x$ axis, the system is invariant under the mirror reflection in the $yz$ plane (\textit{i.e.}, $\mathcal{M}_{x}:x\rightarrow -x$) --and so are $\sigma _{ijk}$-- whereby the nonlinear current density and electric field follow the change $\{ E_{x};j^{(2)}_{x},j^{(2)}_{y}\} \rightarrow $ $\{-E_{x};-j^{(2)}_{x},j^{(2)}_{y}\} $. As a result, the quadratic response relation
must satisfy $j^{(2)}_{x}=\sigma _{xxx}E_{x}E_{x}=-j^{(2)}_{x}$, so $\sigma
_{xxx}( \mathbf{m=\pm x}) =0$. Similarly, when $\mathbf{m}$ is parallel to the $y$ axis, the mirror reflection in the $xz$ plane (\textit{i.e.}, $\mathcal{M}_{y}:y\rightarrow -y$) leads to $j^{(2)}_{y}=$ $\sigma_{yxx}E_{x}E_{x}=-j^{(2)}_{y}$, \textit{i.e.}, $\sigma _{yxx}\left(\mathbf{m=\pm y}\right) =0$.

To characterize the nonlinear transport, we introduce the UMR coefficients $\tilde{\zeta}_{\parallel }^{(2)} = \tilde{\zeta}_x^{(2)}$ and $\tilde{\zeta} _{\perp}^{(2)} = \tilde{\zeta}_y^{(2)}$, where \cite{mehraeen2022spin}
\begin{equation}
\label{Eq:zeta}
\tilde{\zeta}_i^{(2)}
\equiv
\frac{\tilde{\sigma}_{ix}(E_{x}) - \tilde{\sigma}_{ix}(-E_{x})}{\sigma_D E_x}
\simeq
-\frac{2 \tilde{\sigma} _{ixx}}{\sigma _{D}}.
\end{equation}
Here $\tilde{\sigma}_{ij}=j_{i}/E_{j}$ denotes the linear conductivity tensor, and $\sigma _{D}$ is the Drude conductivity of the NM layer. In this definition, the UMR coefficient is solely a property of the system and is independent of the strength of the applied current. Since it has the dimensions of inverse electric field, physically, one can say that it sets the scale of the electric field for which the magnitudes of the UMRs become comparable to linear-response effects \footnote{Note, however, that--owing to their directional nature--UMRs are typically observed at much weaker electric fields, as one need simply reverse the direction of the applied current to detect them.}.

\begin{figure}[tph]
    \sidesubfloat[]{\includegraphics[width=0.75\linewidth,trim={1.5cm 0.7cm 0.5cm 0}]{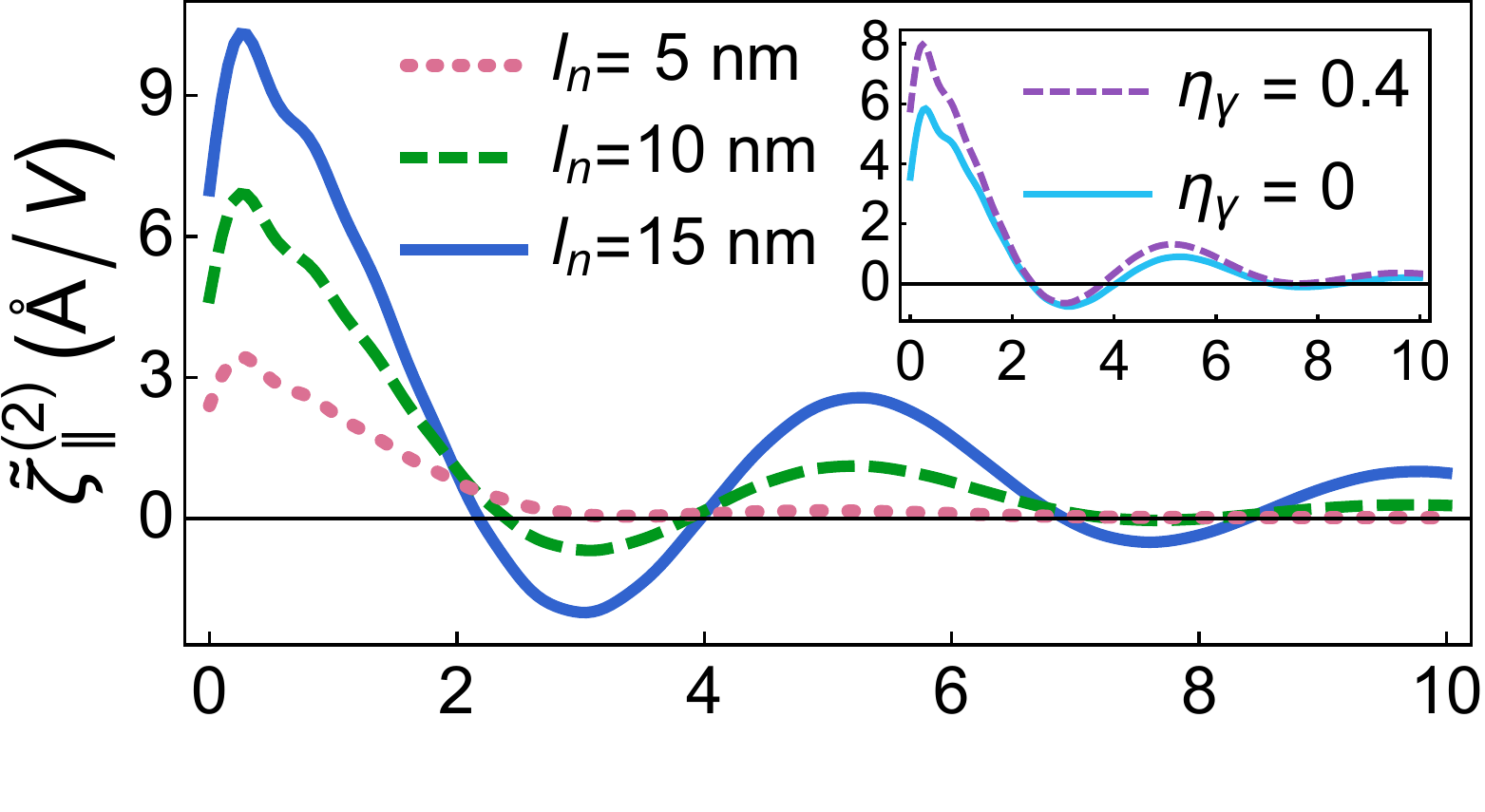}\label{fig4a}}
    \\
    \sidesubfloat[]{\includegraphics[width=0.75\linewidth,trim={1.5cm 0.5cm 0.5cm 0}]{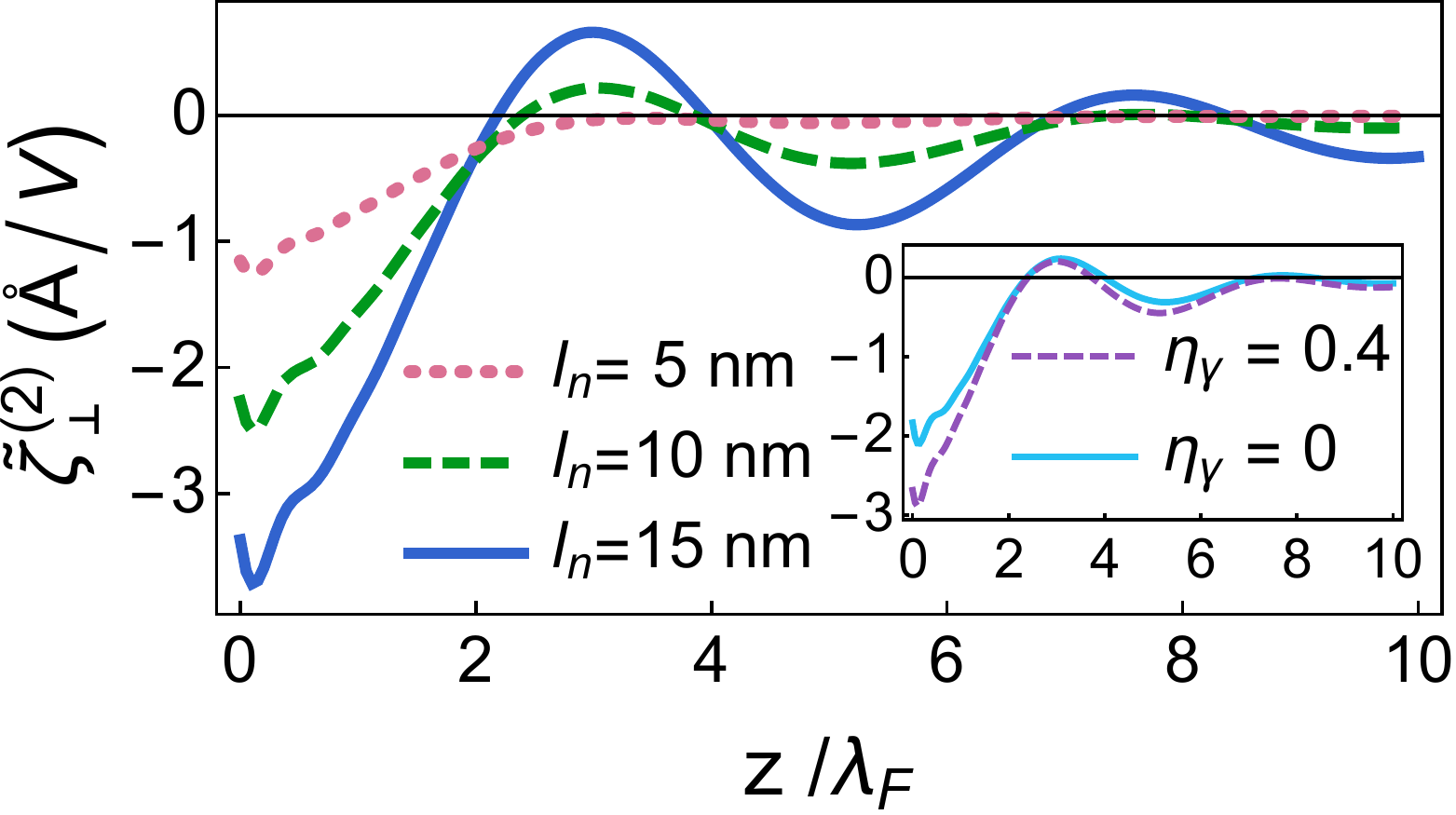}\label{fig4b}}%
    \caption{Plots of the z dependences of the a) longitudinal (at $\mathbf{m}=\mathbf{y}$) and b) transverse (at $\mathbf{m}=\mathbf{x}$) disorder-averaged UMR coefficients for various values of the mean free path, with $\eta_{\gamma}= 0.2$ (main) and for different impurity strengths, with $l_n=10$ nm (inset). Other parameters used: $\eta_R = 0.2$ \cite{mihai2010current,park2013orbital,tokatly2015current,grytsyuk2016k}, $\eta_{ex} = 0.2$ \cite{kajiwara2010transmission}, $m=1.8 \times 10^{-30}$ kg, $\epsilon_F=5$ eV and $V_b=8$ eV.}
    \label{fig4}
\end{figure}

Plots of the $z$ dependencies of the UMR coefficients for various values of the bulk and interfacial disorder parameters $l_n$ and $\eta_{\gamma}$ are presented in Fig.~\ref{fig4}, which reveal that the UMR coefficients scale linearly with the mean free path. This is not surprising, as in the semiclassical picture, $\sigma_{ijk} \propto l_n^2$, while $\sigma_D \propto l_n$. Furthermore, as shown in the insets of Fig.~\ref{fig4}, introducing the interfacial disorder leads to a slight enhancement of the UMR coefficients, which is due to the momentum relaxation of electrons caused by the interfacial disorder that effectively modulates the momentum-dependent spin-orbit scattering [see Eq.~(\ref{R_tilde})]. However, this trend is expected to be reversed when the interfacial disorder is increased further, as the vertex corrections, which constitute the diffuse scattering and contribute negatively to the magnitudes of the UMR coefficients (see Appendix~\ref{appendixB} for details), will play an increasingly important role in the quantum transport.

To shed light on the physical origin of the UMR effect, we first note from Fig.~\ref{fig4} that spatial variations in $\tilde{\zeta}_{\parallel }^{(2)}$ and $\tilde{\zeta}_{\perp }^{(2)}$--comprised of an oscillatory exponential decay--occur over a length scale given by the Fermi wavelength $\lambda _{F}=2\pi/k_F$, with $k_{F}=\sqrt{2m \epsilon _{F}/\hbar ^{2}}$ the Fermi wavevector in the NM layer. This reflects the quantum nature of the nonlinear transport effect, as semiclassical UMRs typically scale with the spin diffusion length. Furthermore, as shown in Appendix~\ref{appendixB}, in the limiting case of $l_n\gg\lambda_F$, Eqs.~(\ref{eq:sigma_ijk}) may be reexpressed entirely in terms of the dressed interference velocity $\braket{\hat{\tilde{\bs{v}}}({\mathbf{q}},z)}_{I-R}$ as
\begin{equation}
\label{sigma_ijk_ballistic}
\tilde{\sigma}_{ijk} ( z ; \mathbf{m})
=
\frac{2 e^3 m^{2}}{\pi \hbar^4 k_F^2} l_n^2
\int \limits_{\mathbf{q}}
\mathcal{F}_{jk}(\mathbf{q})
\Braket{\hat{\tilde{v}}_{i}(\mathbf{q},z)}_{I-R},
\end{equation}
where
\begin{equation}
\mathcal{F}_{jk}(\mathbf{q})
=
\frac{1}{k_{z,F}} 
\left[
q_j q_k
\left(\frac{2}{k_{z,F}^2} + \frac{1}{k_F^2} - \frac{\hbar^2}{m}\pd_{\epsilon_F}\right)
+
\delta_{jk}
\right],
\end{equation}
thereby confirming the quantum-interference origin of the nonlinear magnetoresistances. In Eq.~(\ref{sigma_ijk_ballistic}), we have explicitly noted the dependence of $\tilde{\sigma}_{ijk}$ on the spatial coordinate $z$ and the magnetization $\mathbf{m}$, a property inherited from the interference velocity. Note also that the nonlinear conductivity tensor scales quadratically with the bulk disorder parameter, $\tilde{\sigma}_{ijk} \propto l_n^2$, in agreement with semiclassical expectations.

Incidentally, Eq.~(\ref{sigma_ijk_ballistic}) also reveals a simple relation between the longitudinal and transverse QUMRs. A comparison of the plots in Fig.~\ref{fig4} suggests that the transverse and longitudinal UMR coefficients are related by a factor of -1/3. To confirm this, we note that the point of difference between $\tilde{\sigma}_{xxx}$ and $\tilde{\sigma}_{yxx}$ in Eq.~(\ref{sigma_ijk_ballistic}) is in the azimuthal integrals in momentum space. Performing these, we find that 
$\tilde{\sigma}_{xxx}
\propto
m_y \int_0^{2\pi} d \phi_{\mathbf{q}} \cos^4 \phi_{\mathbf{q}}
=
(3\pi/4) m_y$, while 
$\tilde{\sigma}_{yxx}
\propto
-m_x \int_0^{2\pi} d \phi_{\mathbf{q}} \cos^2 \phi_{\mathbf{q}} \sin^2 \phi_{\mathbf{q}}
=
-(\pi/4) m_x$, with $\phi_{\mathbf{q}}$ the azimuthal angle in momentum space. Thus, for an arbitrary orientation of the magnetization, this gives the simple-- yet general--relation
\begin{equation}
\label{ratio}
\frac{\tilde{\zeta}_{\perp}^{(2)}}
{\tilde{\zeta}_{\parallel}^{(2)}}
=
-\frac{1}{3} \cot \theta_{\mathbf{m}},
\end{equation}
with $\theta_{\mathbf{m}}$ the angle between the projected magnetization onto the layer plane and the applied electric field (see Fig.~\ref{fig1}). And comparison of the transverse UMR coefficient when $\mathbf{m}=\mathbf{x}$ and the longitudinal UMR coefficient when $\mathbf{m} = \mathbf{y}$ readily yields the ratio -1/3 observed in Fig.~\ref{fig4}. This simple relation reflects the common physical origin of the longitudinal and transverse QUMRs, and may serve as an additional transport signature of the quantum nonlinear transport effect.

\section{Materials considerations}

The most promising materials systems to
observe the QUMR effect are probably bilayers consisting of a heavy metal
and a FI, such as Au$|$YIG (Yttrium iron garnet) and $\beta $-Ta$|$YIG.
These systems possess a magnetic interface with sizable Rashba SOC and exchange interaction--two essential ingredients for generating the QUMR.

In principle, the QUMR may also arise in metallic bilayers comprising a
heavy metal and a ferromagnetic metal, but it would be accompanied by other nonlinear effects. One main competitor is the spin-Hall UMR~\cite{avci2015natphys,shulei2016prb,avci2018prl}, arising from the spin accumulation built in the
ferromagnetic-metal layer due to spin-current injection driven by the spin
Hall effect~\cite{dyakonov1971,hirsch1999prl,shufeng2000prl} in the heavy-metal layer. For a typical metallic bilayer,
such as Pt$|$Co, the corresponding UMR coefficient is about $\zeta
^{(2)}\sim 1$ \AA/V, about the same order of magnitude as the predicted
QUMR in a typical NM$|$FI bilayer. Other nonlinear effects that may
intertwine with the QUMR include the anomalous Nernst effect~\cite{syHuang11PRL_ANE} and spin Seebeck effect~\cite{uchida2008observation, uchida2010observation, qu2013prl, adachi2013theory}, arising from a vertical temperature gradient $\boldsymbol{\nabla }T\left(
\propto \mathbf{E}^{2}\right) $ across the ferromagnetic layer induced by Joule heating. Extra care would thus be needed to separate these contributions to the nonlinear resistance from the QUMR in metallic magnetic bilayers~\cite{avci2014interplay}.\\


\section{Closing Remarks}

In conclusion, we have explored the quantum transport of electrons in the nonlinear response regime, and predicted a QUMR effect in NM$|$FI bilayers, which originates from the interference between electron waves approaching and reflecting off a magnetic interface with Rashba SOC. 

Several appealing properties of the QUMR have been identified, enabling both electric and magnetic control of the nonlinear magnetotransport effect. Significantly, the emergence of the QUMR effect entails variations in both longitudinal and transverse resistances whenever the direction of the electric field is reversed. Moreover, the QUMR is also sensitive to the orientation of magnetization of the FI layer, as the phase difference between the scattering waves can be tuned by the magnetization through its coupling with the spin angular momenta of conduction electrons in the NM layer.

As a final remark, we have restricted ourselves to zero temperature in the initial study of the QUMR effect. To examine finite temperature effects, one needs to consider the influence of fluctuating
scatterers on the phase coherence of the quantum magnetotransport in the nonlinear response regime, which warrants future theoretical and experimental research. And from an applications perspective, we envision this work will open new avenues for developing future quantum spintronic devices.

\section*{Acknowledgments}

The authors are grateful to Giovanni Vignale for helpful comments and for critical reading of the manuscript. This work made use of the High Performance Computing Resource in the Core Facility for Advanced Research Computing at Case Western Reserve University. This work was supported by the College of Arts and Sciences, Case Western Reserve University. 

\appendix
\counterwithin{figure}{section}
\onecolumngrid

\section{Scattering Amplitude Matrices}
\label{appendixA}

In this section, we derive the scattering matrix through two equivalent methods. The first approach involves imposing appropriate boundary conditions directly on the wavefunction and its spatial derivative. The main advantage of this approach is that the scattering amplitudes may be obtained in a rather straightforward manner. The second method requires the single-particle electron propagator, which, in turn, allows for a relatively transparent generalization to the case with disorder. This generalization is the focus of the last part of this section.

\subsection{Reflection Matrix from Wavefunction}

Consider a 3D electron gas in the scattering potential \cite{Stiles13PRB_SOT,tokatly2015prb}
\begin{equation}
\hat{V}_{\text{scat.}}(z)=\frac{\alpha_{R}}{\hbar} \delta(z) \hat{\boldsymbol{\sigma}}
\cdot(\hat{\mathbf{p}} \times \mathbf{z})-J_{e x} \delta(z) \hat{\boldsymbol{%
\sigma}} \cdot \mathbf{m} +V_{b} \Theta(-z)\,,
\end{equation}
where $\alpha _{R}$ and $J_{ex}$ are the coefficients of the interfacial
Rashba spin-orbit coupling (SOC) and exchange interaction, respectively, and
$V_{b}$ is the height of the energy barrier, which is greater than the Fermi
energy of electrons in the NM layer.

The general scattering state, given by Eq.~(\ref{gen_scat}), reads
\begin{equation}
\boldsymbol{\psi }_{\text{scat.}}=\left\{
\begin{array}{c}
e^{i\mathbf{q}\cdot \rho }\left( e^{-ik_{z}z}+e^{ik_{z}z}\hat{R}_{\mathbf{q}}\right)
\boldsymbol{\chi }\,,\;\;z>0\;\;~(\text{NM Layer}) \\
~~~~~~e^{i\mathbf{q}\cdot \rho }e^{\kappa_z z}\;\hat{T}_{\mathbf{q}} \boldsymbol{\chi }%
\,,\;\;~~~~~~~~~~z<0\;\;~(\text{FI Layer})%
\end{array}%
,\right.
\end{equation}%
\noindent where $\mathbf{q}\left[ =\left( k_{x},k_{y}\right) \right] $ and $%
\boldsymbol{\rho }\left[ =\left( x,y\right) \right] $ are the wave- and
position-vectors in the $x$-$y$ plane, wherein the system is translationally
invariant and allows propagation of plane waves, $k_{z}$ is the $z$%
-component of the wavevector of the propagating wave in the NM layer, and $%
\kappa_z^{-1}\left[ =\left( 2m V_{b}/\hbar ^{2}-k_{z}^{2}\right)
^{-1/2}\right] $ characterizes the decay length of the evanescent wave in
the FI layer. $\hat{R}_{\mathbf{q}}$ and $\hat{T}_{\mathbf{q}}$ are $2\times 2$ matrices in spin
space, which describe, respectively, the spin-dependent reflection and
transmission amplitudes, and the spinors $\boldsymbol{\chi }$ can be taken
as an arbitrary superposition of the eigenspinors that satisfy $\hat{%
\boldsymbol{\sigma }}\cdot \mathbf{m}\;\chi ^{\pm }=\pm \chi ^{\pm }$.

\begin{figure}[tph]
    \sidesubfloat[]{\includegraphics[width=0.25\linewidth,trim={1.5cm 0 0cm 0}]{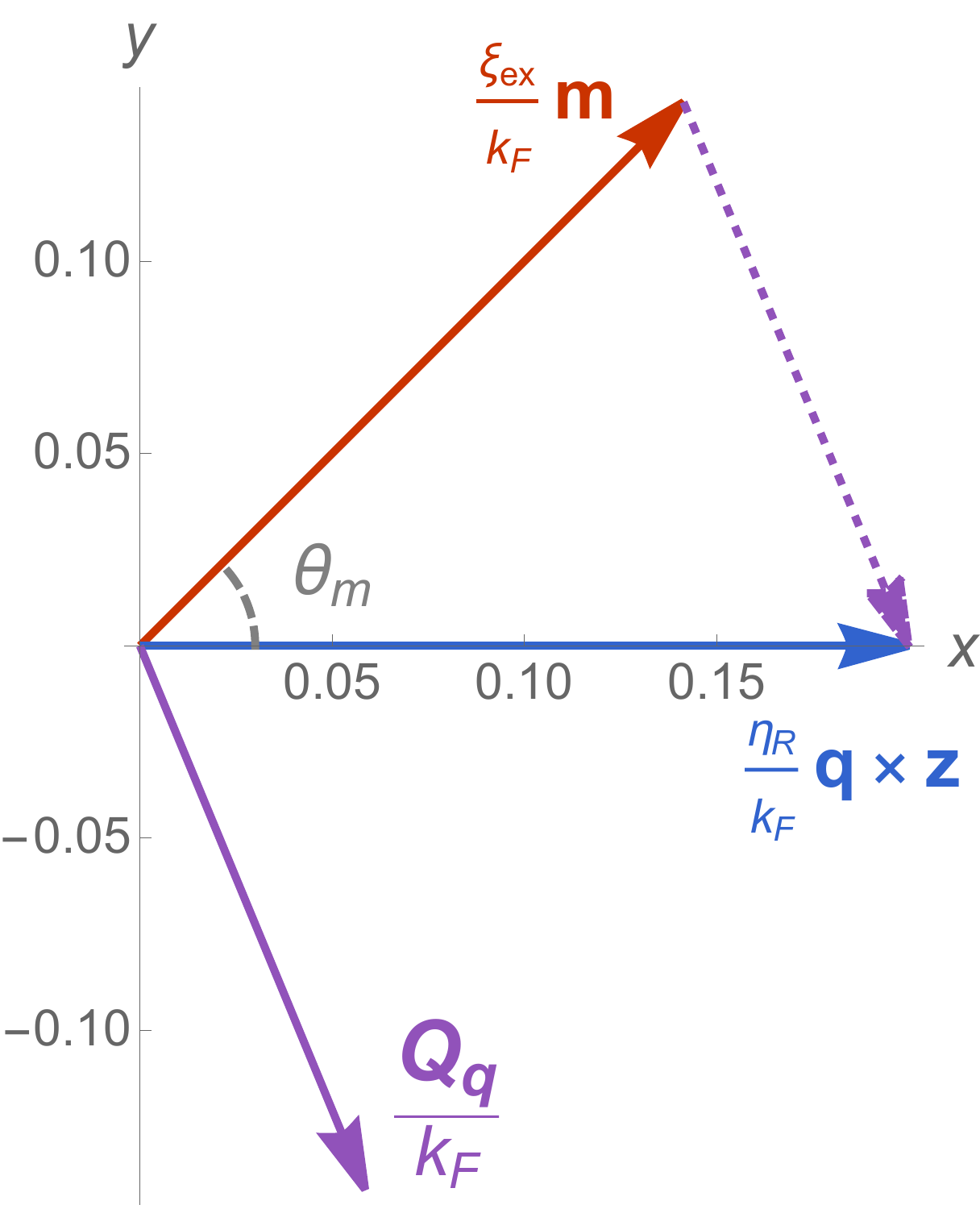}\label{figS1a}}\quad%
    \sidesubfloat[]{\includegraphics[width=0.25\linewidth,trim={1.5cm 1cm 0.5cm 0}]{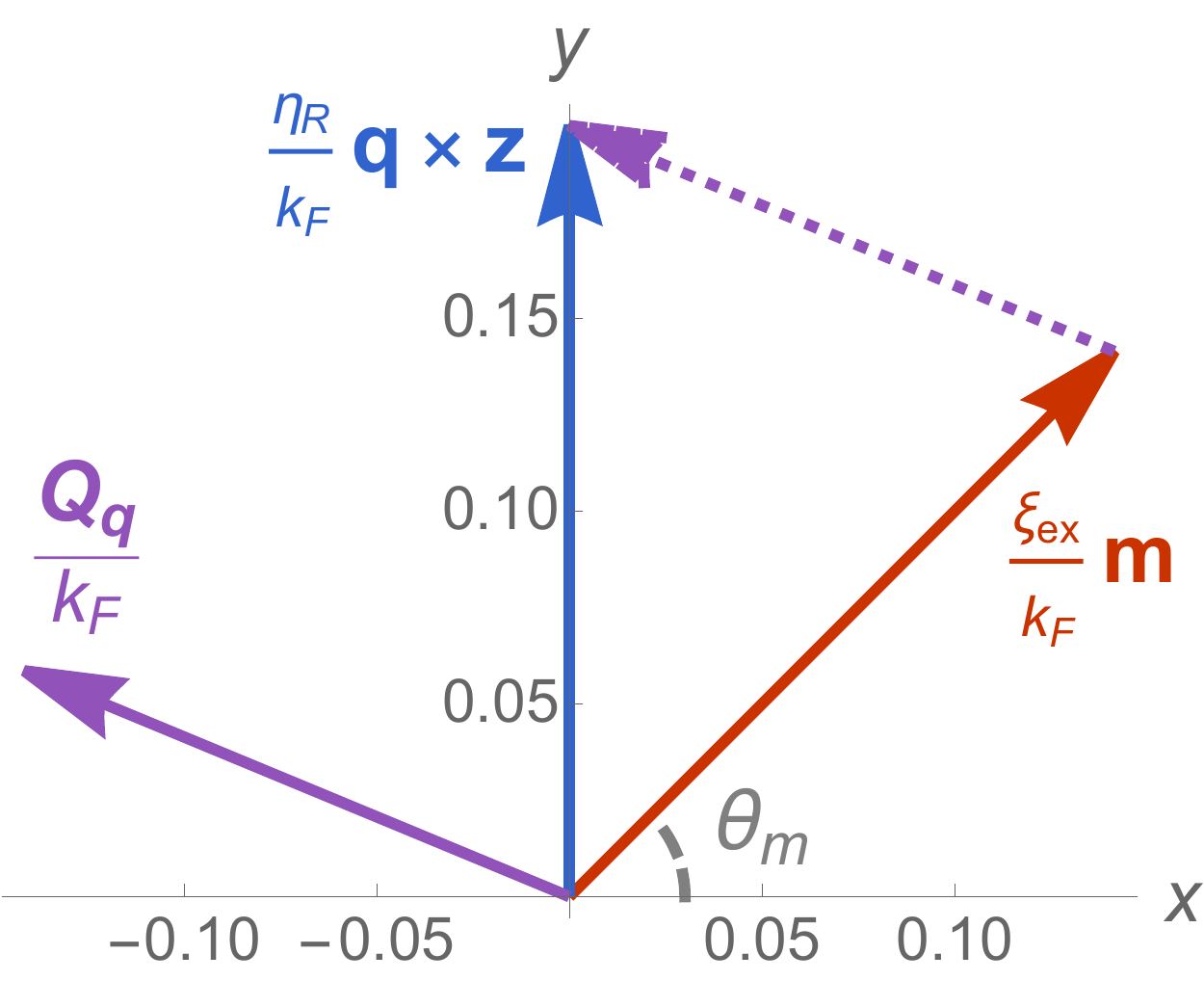}\label{figS1b}}%
    \caption{Plots of the dimensionless vectors $\eta_R \mathbf{q}\times \mathbf{z}/k_F$, $\xi_{ex}\mathbf{m}/k_F$ and $\mathbf{Q_q}/k_F=(\eta_R \mathbf{q}\times \mathbf{z} - \xi_{ex}\mathbf{m})/k_F$ for the unit vectors (a) $\mathbf{q}\times \mathbf{z}/k_F=\mathbf{x}$ and (b) $\mathbf{q}\times \mathbf{z}/k_F=\mathbf{y}$. Parameters used: $\eta_R=\xi_{\text{ex}}/k_F=0.2$ and $\theta_m=\cos^{-1}(\mathbf{m}\cdot \mathbf{x})=\pi/4$.}
    \label{figS1}
\end{figure}

The wavefunction obeys the following boundary conditions at the Rashba
interface
\begin{subequations}
\begin{gather}
\boldsymbol{\psi }_{\text{scat.}}\left( \rho ,0^{+}\right) =\boldsymbol{\psi
}_{\text{scat.}}\left( \rho ,0^{-}\right),
\label{BC1}
\\
\frac{d}{dz}\boldsymbol{\psi }_{\text{scat.}}\left( \rho ,0^{+}\right) -%
\frac{d}{dz}\boldsymbol{\psi }_{\text{scat.}}\left( \rho ,0^{-}\right) =\hat{%
\boldsymbol{\sigma }}\cdot \mathbf{Q_{q}}\boldsymbol{\psi }_{\text{scat.}%
}(\rho ,0),
\label{BC2}
\end{gather}
\end{subequations}
where $\mathbf{Q_{q}}=\eta _{R}\mathbf{q}\times \mathbf{z}-\xi _{ex}\mathbf{m%
}$ --as shown in Fig.~\ref{figS1}-- with $\eta _{R}\equiv 2m \alpha _{R}/\hbar ^{2}$ a
dimensionless constant characterizing the strength of the interfacial Rashba
SOC and $\xi _{ex}\equiv 2m J_{ex}/\hbar ^{2}$ the rescaled
exchange interaction. Eliminating $\boldsymbol{\chi }$ and $e^{i\mathbf{q}%
\cdot \rho }$ from Eqs.~(\ref{BC1}) and~(\ref{BC2}), we obtain a set of
equations for the scattering amplitude matrices:
\begin{subequations}
\begin{gather}
1+\hat{R}_{\mathbf{q}}=\hat{T}_{\mathbf{q}},
\\
-ik_{z}\left( 1-\hat{R}_{\mathbf{q}}\right) -\kappa_z \hat{T}_{\mathbf{q}}=\hat{\boldsymbol{\sigma }}%
\cdot \mathbf{Q_{q}}\hat{T}_{\mathbf{q}}.
\end{gather}
\end{subequations}

\noindent Solving the set of equations, one can obtain the general expression of the
reflection amplitude matrix as follows
\begin{equation}
\hat{R}_{\mathbf{q}}=\frac{\mathbf{Q}_{\mathbf{q}}^{2}-(\kappa_z ^{2}+k_{z}^{2})+2ik_{z}%
\hat{\boldsymbol{\sigma }}\cdot \mathbf{Q}_{\mathbf{q}}}{(\kappa_z
-ik_{z})^{2}-\mathbf{Q}_{\mathbf{q}}^{2}}. \label{rmatrix}
\end{equation}
\noindent Introducing the change of variables $\varphi _{\mathbf{q}}=\arcsin \left( 2k_{z}\kappa_z /\varkappa_{\mathbf{q}}\right)$ and $\vartheta _{\mathbf{q}}=\arcsin \left( 2k_{z}Q_{\mathbf{q}}/\varkappa_{\mathbf{q}}\right)$ with $\varkappa_{\mathbf{q}}^{2}=\left[ Q_{%
\mathbf{q}}^{2}-(\kappa_z ^{2}+k_{z}^{2})\right] ^{2}+\left( 2k_{z}Q_{\mathbf{q%
}}\right) ^{2}$, and using the unit vector $\mathbf{n}_{\mathbf{q}} \mathbf{=\mathbf{Q}_{%
\mathbf{q}}/}Q_{\mathbf{q}}$, the reflection matrix is recast in the form
\begin{equation}
\label{SEq:R_q}
\hat{R}_{\mathbf{q}}
=
e^{i \left( \varphi_{\mathbf{q}} + \vartheta _{\mathbf{q}}
\hat{\boldsymbol{\sigma}}\cdot \mathbf{n}_{\mathbf{q}} \right)},
\end{equation}
which is Eq.~(\ref{eq:R_q}) presented in the main text.

\subsection{Reflection Matrix from Propagator}

\subsubsection{Propagator without Interfacial Effects}

Let us rewrite the Hamiltonian given by Eq.~(\ref{Ham}) as
\begin{subequations}
\label{Seq:H_hat}
\begin{gather}
\hat{H}
=
H^0 + \hat{V}^{\text{int}}_{\mathbf{q}} \delta(z),
\\
H^0
=
\frac{\hbar^2 k^2}{2m} + V_b \Theta(-z),
\\
\hat{V}^{\text{int}}_{\mathbf{q}}
=
\alpha_R \,\hat{\bs{\sigma}} \cdot \mathbf{q} \times \mathbf{z} - J_{ex}\hat{\bs{\sigma}} \cdot \mathbf{m},
\end{gather}
\end{subequations}
where $\mathbf{q}$ is the in-plane momentum and $\hat{V}^{\text{int}}_{\mathbf{q}}$ is the interaction potential at the interface and consists of the interfacial Rashba SOC and exchange interactions. For the unperturbed system, $H^0$, the general scattering state for electrons with momentum $\mathbf{k}=(\mathbf{q},k_1)$ incident on the interface from the NM reads
\begin{equation}
\label{SEq:psi_k}
\psi_{\mathbf{k}}\left(\mathbf{r}\right)
=
\phi_{k_1/\kappa}(z) e^{i \mathbf{q} \cdot \bs{\rho}},
\end{equation}
where
\begin{equation}
\phi_{k_1/\kappa}(z)
=
\begin{cases}
      e^{-i k_1 z} + a_r \left( k_1, \kappa \right) e^{i k_1 z}, & z>0\; \; \text{(NM)}  \\
      a_t \left( k_1, \kappa \right) e^{\kappa z}, & z<0 \; \; \text{(FI)}  
    \end{cases}.
\end{equation}
Here, $\kappa$ quantifies the decay rate of the wavefunction in the FI and $\mathbf{r}=(\bs{\rho},z)$ is the position vector, with $a_r$ and $a_t$ the reflection and transmission amplitudes, respectively. These may be obtained from the boundary conditions on the wave function
\begin{subequations}
\begin{gather}
\phi_{k_1}\left(0^+\right)
=
\phi_{\kappa}\left(0^-\right),
\\
\frac{d}{dz}\phi_{k_1}\left(0^+\right)
=
\frac{d}{dz}\phi_{\kappa}\left(0^-\right),
\end{gather}
\end{subequations}
which imply 
\begin{equation}
a_r \left( k_1, \kappa \right)
=
\frac{ik_1 + \kappa}{ik_1 - \kappa},
\end{equation}
and $a_t=1+a_r$. The retarded unperturbed Green's function is
\begin{equation}
G^{0,R} \left(\epsilon\right)
=
\left(\epsilon - H^0 + i\delta \right)^{-1},
\end{equation}
which gives rise to the real-space propagator
\begin{equation}
\label{SEq:G_0,R}
G^{0,R} \left(\mathbf{r},\mathbf{r}^{\prime};\epsilon\right)
\equiv
\Braket{\mathbf{r} | G^{0,R} \left(\epsilon\right) | \mathbf{r}^{\prime}}
=
\sum_{\mathbf{k}} 
\frac{\psi_{\mathbf{k}}\left(\mathbf{r}\right) \psi_{\mathbf{k}}^*\left(\mathbf{r}^{\prime}\right)}
{\epsilon - \epsilon_{\mathbf{k}} + i\delta},
\end{equation}
where $\epsilon_{\mathbf{k}}= \hbar^2 k^2 /2 m + V_b \Theta(-z)$. Inserting Eq.~(\ref{SEq:psi_k}) into Eq.~(\ref{SEq:G_0,R}), we obtain the decomposition
\begin{equation}
\label{gR_z}
G^{0,R} \left(\mathbf{r},\mathbf{r}^{\prime};\epsilon\right)
=
\sum_{\mathbf{q}} e^{i\mathbf{q} \cdot(\bs{\rho} - \bs{\rho}^{\prime})} g^{0,R}_{\mathbf{q}}\left(z,z^{\prime};\epsilon\right),
\end{equation}
which, for correlations in the NM layer, leads to the solution
\begin{equation}
\label{gR_0_nm}
g^{0,R}_{\mathbf{q}}\left(z>0,z^{\prime}>0;\epsilon\right)
=
\sum_{k_1} 
\frac{\phi_{k_1}\left(z\right) \phi_{k_1}^*\left(z^{\prime}\right)}
{\epsilon - \epsilon_{\mathbf{k}} + i\delta}
=
\frac{m}{i \hbar^2 k_z(\epsilon)} \left[e^{i k_z(\epsilon) |z-z^{\prime}|} + a_r \left( k_z, \kappa_z \right) e^{i k_z(\epsilon)\left(z+z^{\prime}\right)}\right].
\end{equation}
Here, $k_1= \pm k_z(\epsilon)$ are the poles of the propagator at energy $\epsilon$, with $k_z(\epsilon)=\sqrt{[k(\epsilon)]^2 - q^2}$, $k(\epsilon)=\sqrt{2m |\epsilon| /\hbar^2}$, $\kappa_z(\epsilon)=\sqrt{k_b^2 - [k_z(\epsilon)]^2}$ and $k_b=\sqrt{2m V_b/\hbar^2}$.

Before proceeding to include interfacial effects, we also add the contribution of the bulk scatterers to the propagator. This can be achieved by modifying the Hamiltonian as $H^0 \rightarrow H^0 + \Sigma ^B(\mathbf{r})$, where $\Sigma^B(\mathbf{r})$ is the local self-energy due to bulk impurity scattering, resulting in a local scattering time  $\tau(\mathbf{r})= - \hbar/2 \text{Im} [\Sigma^B(\mathbf{r})]$ and a reduction of the propagator as \cite{camblong1994theory}
\begin{equation}
G^{0,R} \left(\mathbf{r},\mathbf{r}^{\prime};\epsilon\right)
\rightarrow
G^{0,R} \left(\mathbf{r},\mathbf{r}^{\prime};\epsilon\right) \text{exp} \left[- \int \limits_{\Gamma\left[\mathbf{r}, \mathbf{r}^{\prime}\right]} \frac{ds^{\dprime}}{2 l\left(\mathbf{r}^{\dprime}\right)} \right],
\end{equation}
which arises as a result of the damping of the electron wavefunction as it propagates along the straight path $\Gamma\left[\mathbf{r}, \mathbf{r}^{\prime}\right]$ connecting the points $\mathbf{r}$ and $\mathbf{r}^{\prime}$, and accounts for the average scattering encountered by the electron. Here, $l(\mathbf{r})=v_F \tau(\mathbf{r})$ is the local scattering length of the conduction electrons. In the NM layer, the loss of momentum associated with the damping of the wavefunction corresponds to the replacement $k_z \rightarrow k_z \sqrt{1 + i k_F/k_z^2 l_n} \simeq k_z + i \frac{k_F}{2 k_z l_n}$ at the Fermi level, where $l_n$ is the mean free path of the conduction electrons and $k_F = \sqrt{2m \epsilon_F /\hbar^2}$ is the Fermi wavenumber. Thus, the unperturbed propagator now reads
\begin{equation}
\label{gR_0_Ln}
g^{0,R}_{\mathbf{q}}\left(z,z^{\prime};\epsilon\right)
=
\frac{m}{i \hbar^2 k_z(\epsilon)} \left[e^{i k_z(\epsilon) |z-z^{\prime}|} e^{- k(\epsilon) |z-z^{\prime}|/2 k_z(\epsilon) l_n} 
+
a_r\left(k_z, \kappa_z\right) e^{i k_z(\epsilon)\left(z+z^{\prime}\right)}e^{- k(\epsilon) \left(z+z^{\prime}\right)/2 k_z(\epsilon) l_n}\right].
\end{equation}

\subsubsection{Including Interfacial Potential}

In the presence of spin-dependent interactions at the interface, the propagator is now a $2\times2$ matrix in spin space, which can be found using the Dyson equation
\begin{equation}
\label{GR_soc}
\hat{G}^R \left(\mathbf{r},\mathbf{r}^{\prime};\epsilon\right)
=
\hat{G}^{0,R} \left(\mathbf{r},\mathbf{r}^{\prime};\epsilon\right)
+
\int d\mathbf{r}_1 \hat{G}^{0,R} \left(\mathbf{r},\mathbf{r}_1;\epsilon\right) \hat{V}^{\text{int}}_{\mathbf{q}} \delta(z_1) \hat{G}^R \left(\mathbf{r}_1,\mathbf{r}^{\prime};\epsilon\right)\,,
\end{equation}
where
\begin{equation}
\hat{G}^{0,R} \left(\mathbf{r},\mathbf{r}^{\prime};\epsilon\right)
=
G^{0,R} \left(\mathbf{r},\mathbf{r}^{\prime};\epsilon\right) \hat{\sigma}_0\,.
\end{equation}
Inserting Eq.~(\ref{gR_z}) into Eq.~(\ref{GR_soc}), and introducing the analogous decomposition
\begin{equation}
\label{gR_hat_z}
\hat{G}^R \left(\mathbf{r},\mathbf{r}^{\prime};\epsilon\right)
=
\sum_{\mathbf{q}} e^{i\mathbf{q} \cdot(\bs{\rho} - \bs{\rho}^{\prime})} \hat{g}^R_{\mathbf{q}} \left(z,z^{\prime};\epsilon\right),
\end{equation}
the Fourier-transformed Dyson equation reads
\begin{equation}
\hat{g}^R_{\mathbf{q}}\left(z,z^{\prime};\epsilon\right)
=
\hat{g}^{0,R}_{\mathbf{q}}\left(z,z^{\prime};\epsilon\right)
+
\hat{g}^{0,R}_{\mathbf{q}}\left(z,0;\epsilon\right) \hat{V}^{\text{int}}_{\mathbf{q}}
\hat{g}^R_{\mathbf{q}}\left(0,z^{\prime};\epsilon\right),
\end{equation}
so that for interfacial correlations ($z,z^{\prime}=0$), the decomposed propagator is obtained as
\begin{equation}
\label{gR_hat_0}
\hat{g}^R_{\mathbf{q}}
=
\left(\left[\hat{g}^{0,R}_{\mathbf{q}}\right]^{-1} - \hat{V}_{\mathbf{q}}\right)^{-1}
=
\frac{2m}{\hbar^2} \left[\left(i k_z - \kappa_z \right) \hat{\sigma}_0 - \hat{\bs{\sigma}} \cdot \mathbf{Q}_{\mathbf{q}}\right]^{-1}.
\end{equation}
On the other hand, spin-dependent scattering at the interface results in spin-dependent reflection of the electron wavefunction, so that Eq.~(\ref{gR_0_Ln}) may be generalized to
\begin{equation}
\label{gR_hat_nm}
\hat{g}^R_{\mathbf{q}} \left(z,z^{\prime};\epsilon\right)
=
\frac{m}{i \hbar^2 k_z(\epsilon)} \left[e^{i k_z(\epsilon) |z-z^{\prime}|}e^{- k(\epsilon) |z-z^{\prime}|/2 k_z(\epsilon) l_n}
+
\hat{R}_{\mathbf{q}} e^{i k_z(\epsilon)\left(z+z^{\prime}\right)}e^{- k(\epsilon) \left(z+z^{\prime}\right)/2 k_z(\epsilon) l_n}\right],
\end{equation}
where $\hat{R}_{\mathbf{q}}$ is the reflection matrix. Equating Eq.~(\ref{gR_hat_nm}) at $z,z^{\prime}=0$ to Eq.~(\ref{gR_hat_0}), we arrive at the general form of the reflection matrix
\begin{equation}
\label{Seq:R_hat1}
\hat{R}_{\mathbf{q}}
=
\frac{\mathbf{Q}_{\mathbf{q}}^2 -\left(k_z^2 + \kappa_z^2 \right) + 2ik_z \hat{\bs{\sigma}} \cdot \mathbf{Q}_{\mathbf{q}}}
{\left(ik_z - \kappa_z \right)^2 - \mathbf{Q}_{\mathbf{q}}^2},
\end{equation}
or, equivalently,
\begin{equation}
\label{Seq:R_hat2}
\hat{R}_{\mathbf{q}}
=
e^{i \left( \varphi_{\mathbf{q}} + \vartheta _{\mathbf{q}}
\hat{\boldsymbol{\sigma}}\cdot \mathbf{n}_{\mathbf{q}} \right)},
\end{equation}
in agreement with Eq.~(\ref{SEq:R_q}), which was derived directly from the wave function.

\subsection{Including Interfacial Disorder}

Having derived the reflection matrix from the Green's function in the absence of interfacial disorder, we now proceed to calculate the disordered propagator by taking into account interfacial impurities, which, in general, lead to both specular and diffuse scattering off the bilayer interface. This calculation is done in two equivalent ways: once from the transition matrix, and then from a self-energy calculation. From the latter approach, we then derive the general form of the disordered reflection matrix. 

\subsubsection{Dressed Propagator from Transition Matrix}

Recall the Dyson equation given by Eq.~(\ref{G_tilde}) in the main text
\begin{equation}
\label{GR_imp}
\hat{\tilde{G}}^R \left(\mathbf{r},\mathbf{r}^{\prime};\epsilon\right)
=
\hat{G}^R\left(\mathbf{r},\mathbf{r}^{\prime};\epsilon\right)
+
\int d\mathbf{r}_1 \hat{G}^R \left(\mathbf{r},\mathbf{r}_1;\epsilon\right) \hat{V}^{\text{imp}}\left({\mathbf{r}}_1\right)\hat{\tilde{G}}^R \left(\mathbf{r}_1,\mathbf{r}^{\prime};\epsilon\right),
\end{equation}
where $\hat{\tilde{G}}^R \left(\mathbf{r},\mathbf{r}^{\prime};\epsilon\right)$ is the dressed propagator in the presence of both disorder and interfacial interaction, and $\hat{V}^{\text{imp}}(\mathbf{r})=\hat{V}^{\text{imp}} (\bs{\rho})\delta(z)$ is the interfacial impurity potential, for which we assume the white noise distribution $\Braket{V^{\text{imp}} \left(\bs{\rho}\right)}
= 0$ and $\Braket{V^{\text{imp}} \left(\bs{\rho}\right) V^{\text{imp}} \left(\bs{\rho}^{\prime}\right)} = \gamma^2 \delta\left(\bs{\rho} - \bs{\rho}^{\prime}\right)$ \cite{stewart2003interfacial}, where the parameter $\gamma$, or its dimensionless counterpart $\eta_{\gamma} = (2m \gamma/\hbar^2)^2$, characterizes the strength of the impurity interaction and $\braket{\cdots}$ here denotes the configurational average over impurity positions.

Disorder at the interface destroys the in-plane periodicity of the system so that the propagator is no longer diagonal in momentum space and is instead decomposed as
\begin{equation}
\label{gR_imp_z}
\hat{\tilde{G}}^R \left(\mathbf{r},\mathbf{r}^{\prime};\epsilon\right)
=
\sum_{\mathbf{q},\mathbf{q}^{\prime}} e^{i \left(\mathbf{q} \cdot\bs{\rho} - \mathbf{q} ^{\prime} \cdot \bs{\rho}^{\prime} \right)}
\hat{\tilde{g}}^R_{\mathbf{q}\mathbf{q}^{\prime}} \left(z,z^{\prime};\epsilon\right).
\end{equation}
Inserting Eq.~(\ref{gR_imp_z}) into Eq.~(\ref{GR_imp}), the Dyson equation is decomposed as
\begin{equation}
\label{gR_imp}
\hat{\tilde{g}}^R_{\mathbf{q}\mathbf{q}^{\prime}} \left(z , z^{\prime};\epsilon\right)
=
\hat{g}^R_{\mathbf{q}} \left(z,z^{\prime};\epsilon\right) \delta_{\mathbf{q}\mathbf{q}^{\prime}}
+
\hat{g}^R_{\mathbf{q}} \left(z,0;\epsilon\right) \hat{\mathcal{T}}_{\mathbf{q}\mathbf{q}^{\prime}}\left(\epsilon\right)
\hat{g}^R_{\mathbf{q}^{\prime}} \left(0,z^{\prime};\epsilon\right),
\end{equation}
where the transition matrix $\hat{\mathcal{T}}_{\mathbf{q}\mathbf{q}^{\prime}}$ obeys the integral equation
\begin{equation}
\label{T_qq}
\hat{\mathcal{T}}_{\mathbf{q}\mathbf{q}^{\prime}}\left(\epsilon\right) 
=
\hat{V}^{\text{imp}}_{\mathbf{q}\mathbf{q}^{\prime}}
+
\sum_{\mathbf{q}_1} \hat{V}^{\text{imp}}_{\mathbf{q}\mathbf{q}_1} \hat{g}^R_{\mathbf{q}_1} \left(\epsilon\right) \hat{\mathcal{T}}_{\mathbf{q}_1\mathbf{q}^{\prime}}\left(\epsilon\right).
\end{equation}
Here, $\hat{g}^R_{\mathbf{q}} \left(\epsilon\right) \equiv \hat{g}^R_{\mathbf{q}} \left(0,0;\epsilon\right)$ and the Fourier transform of the impurity potential is given by
\begin{equation}
\hat{V}^{\text{imp}}_{\mathbf{q}\mathbf{q}^{\prime}}
=
\int d\bs{\rho} \, e^{-i \bs{\rho} \cdot \left(\mathbf{q} - \mathbf{q}^{\prime}\right)} \hat{V}^{\text{imp}}\left(\bs{\rho}\right).
\end{equation}
For completeness, we also note the white noise distribution relations in momentum space
\begin{subequations}
\begin{align}
\Braket{\hat{V}^{\text{imp}}_{\mathbf{q} \mathbf{q}^{\prime}}}
&=
0,
\\
\Braket{\hat{V}^{\text{imp}}_{\mathbf{q} \mathbf{q}^{\prime}} \hat{V}^{\text{imp}}_{\mathbf{q}^{\dprime} \mathbf{q}^{\tprime}}}
&=
\gamma^2 \,\delta_{\mathbf{q} + \mathbf{q}^{\dprime}, \mathbf{q}^{\prime} + \mathbf{q}^{\tprime}}.
\end{align}
\end{subequations}
Upon ensemble averaging, translation invariance is restored and the disorder-averaged propagator becomes diagonal in momentum space,
\begin{equation}
\label{gR_imp_z_avg}
\Braket{\hat{\tilde{G}}^R \left(\mathbf{r},\mathbf{r}^{\prime};\epsilon\right)}
=
\sum_{\mathbf{q}} e^{i \mathbf{q} \cdot \left(\bs{\rho} - \bs{\rho}^{\prime} \right)} \hat{\tilde{g}}^R_{\mathbf{q}} \left(z,z^{\prime};\epsilon\right)\,,
\end{equation}
where we have used the definition $\braket{\hat{\tilde{g}}^R_{\mathbf{q} \mathbf{q}^{\prime}} \left(z,z^{\prime};\epsilon\right)} \equiv \hat{\tilde{g}}^R_{\mathbf{q}} \left(z,z^{\prime};\epsilon\right) \delta_{\mathbf{q}\mathbf{q}^{\prime}}$. Taking the configurational average of Eq.~(\ref{gR_imp}), the dressed propagator reads
\begin{equation}
\hat{\tilde{g}}^R_{\mathbf{q}} \left(z,z^{\prime};\epsilon\right)
=
\hat{g}^R_{\mathbf{q}} \left(z,z^{\prime};\epsilon\right)
+
\hat{g}^R_{\mathbf{q}} \left(z,0;\epsilon\right) \hat{\mathcal{T}}\left( \epsilon \right) \hat{g}^R_{\mathbf{q}} \left(0,z^{\prime};\epsilon\right),
\end{equation}
with the configurational average of the transition matrix defined as $\braket{\hat{\mathcal{T}}_{\mathbf{q}\mathbf{q}^{\prime}}} \equiv \hat{\mathcal{T}} \delta_{\mathbf{q}\mathbf{q}^{\prime}}$, which is found by taking the ensemble average of Eq.~(\ref{T_qq}) and using the Born approximation, leading to the solution
\begin{equation}
\label{T_s}
\hat{\mathcal{T}} \left( \epsilon \right)
=
\gamma^2 \sum_{\mathbf{q}} 
\hat{g}^R_{\mathbf{q}} \left(\epsilon\right).
\end{equation}
Thus, the dressed propagator is obtained. Below, we arrive at the same result from a self-energy calculation, which is slightly more convenient for deriving the dressed reflection matrix.

\subsubsection{Dressed Propagator and Reflection Matrix from Self-Energy}

Consider the configurational average of Eq.~(\ref{GR_imp}),
\begin{equation}
\label{GR_imp_2nd}
\Braket{\hat{\tilde{G}}^R \left(\mathbf{r},\mathbf{r}^{\prime};\epsilon\right)}
=
\hat{G}^R\left(\mathbf{r},\mathbf{r}^{\prime};\epsilon\right)
+
\int d\mathbf{r}_1 \int d\mathbf{r}_2 \,
\hat{G}^R \left(\mathbf{r},\mathbf{r}_1;\epsilon\right) \Braket{\hat{V}^{\text{imp}} \left({\mathbf{r}}_1\right) 
\hat{G}^R \left(\mathbf{r}_1,\mathbf{r}_2;\epsilon\right) \hat{V}^{\text{imp}} \left({\mathbf{r}}_2\right)\hat{\tilde{G}}^R \left(\mathbf{r}_2,\mathbf{r}^{\prime};\epsilon\right)}.
\end{equation}
Using Eq.~(\ref{gR_imp_z_avg}) and applying the expansion \cite{brataas1994semiclassical}
\begin{equation}
\Braket{\hat{V}^{\text{imp}}\left({\mathbf{r}}_1\right)\hat{V}^{\text{imp}}\left({\mathbf{r}}_2\right)\hat{\tilde{G}}_R \left(\mathbf{r}_2,\mathbf{r}^{\prime};\epsilon\right)}
=
\Braket{\hat{V}^{\text{imp}}\left({\mathbf{r}}_1\right)\hat{V}^{\text{imp}}\left({\mathbf{r}}_2\right)}
\Braket{\hat{\tilde{G}}_R \left(\mathbf{r}_2,\mathbf{r}^{\prime};\epsilon\right)}
+
\text{vertex corrections}
,
\end{equation}
Eq.~(\ref{GR_imp_2nd}) is decomposed as
\begin{equation}
\label{gR_imp_sig}
\hat{\tilde{g}}^R_{\mathbf{q}} \left(z,z^{\prime};\epsilon\right)
=
\hat{g}^R_{\mathbf{q}} \left(z,z^{\prime};\epsilon\right)
+
\hat{g}^R_{\mathbf{q}} \left(z,0;\epsilon\right) \hat{\tilde{\Sigma}}^{\text{imp}} \left(\epsilon\right) 
\hat{\tilde{g}}^R_{\mathbf{q}} \left(0,z^{\prime};\epsilon\right),
\end{equation}
where $\hat{\tilde{\Sigma}}^{\text{imp}} \left(\epsilon\right)$ is the configurationally averaged irreducible self-energy and is given, within the Born approximation, by
\begin{equation}
\label{Sigma_tilde}
\hat{\tilde{\Sigma}}^{\text{imp}} \left(\epsilon\right)
=
\gamma^2 \sum_{\mathbf{q}}
\hat{g}^R_{\mathbf{q}} \left(\epsilon\right).
\end{equation}
Note that this is equivalent to the diagonal elements of the disorder-averaged transition matrix, Eq.~(\ref{T_s}), as one would expect \cite{bruus2004many}.

Solving Eq.~(\ref{gR_imp_sig}) for the disordered propagator at $z,z^{\prime}=0$, we find
\begin{equation}
\label{gR_imp_00}
\hat{\tilde{g}}^R
=
\left[\left(\hat{g}^R \right)^{-1} - \hat{\tilde{\Sigma}}^{\text{imp}} \right]^{-1}
=
\left[\left(\hat{g}^{0,R}_{\mathbf{q}} \right)^{-1} -\left(\hat{V}^{\text{int}}_{\mathbf{q}}
+
\hat{\tilde{\Sigma}}^{\text{imp}} \right) \right]^{-1}.
\end{equation}
Thus, in analogy with Eq.~(\ref{gR_hat_nm}), the dressed propagator may be expressed as
\begin{equation}
\label{gR_hat_imp}
\hat{\tilde{g}}^R_{\mathbf{q}}\left(z,z^{\prime};\epsilon\right)
=
\frac{m}{i \hbar^2 k_z(\epsilon)} \left[e^{i k_z(\epsilon) |z-z^{\prime}|}e^{- k(\epsilon) |z-z^{\prime}|/2 k_z(\epsilon) l_n}
+
\hat{\tilde{R}}_{\mathbf{q}} e^{i k_z(\epsilon)\left(z+z^{\prime}\right)}e^{- k(\epsilon) \left(z+z^{\prime}\right)/2 k_z(\epsilon) l_n}\right],
\end{equation}
where, using the decomposition $\hat{\tilde{\Sigma}} \left(\epsilon\right)
=
\tilde{\Sigma}_{\mu}^{\text{imp}} \left(\epsilon\right) \hat{\sigma}^{\mu}$ with $\mu=0,x,y,z$ and introducing the change of variables $\tilde{\Sigma}_{\mu}^{\text{imp}} \left(\epsilon\right) \equiv \hbar^2 \tilde{\xi}_{\mu}\left(\epsilon\right) /2m$, the dressed reflection matrix reads
\begin{equation}
\begin{split}
\label{R_hat_tilde}
\hat{\tilde{R}}_{\mathbf{q}}
&=
\frac{\tilde{\mathbf{Q}}_{\mathbf{q}}^2 -\left(\tilde{\kappa}_z^2 + \tilde{k}_z^2 \right)
\left( 1 + \frac{2i \text{Im} \,\tilde{\xi}_0}
{\tilde{\kappa}_z + i \tilde{k}_z} \right)
+ 
2i k_z \hat{\bs{\sigma}} \cdot \tilde{\mathbf{Q}}_{\mathbf{q}}}
{\left(\tilde{\kappa}_z - i\tilde{k}_z \right)^2 - \tilde{\mathbf{Q}}_{\mathbf{q}}^2},
\end{split}
\end{equation}
with $\tilde{k}_z\left(\epsilon\right) \equiv k_z\left(\epsilon\right) - \text{Im} \;\tilde{\xi}_0\left(\epsilon\right)$, $\tilde{\kappa}_z \left(\epsilon\right) \equiv \kappa_z \left(\epsilon\right) +
\text{Re} \;\tilde{\xi}_0\left(\epsilon\right)$ and $\tilde{\mathbf{Q}}_{\mathbf{q}} \equiv \mathbf{Q}_{\mathbf{q}} + \tilde{\bs{\xi}}\left(\epsilon\right)$. 

Introducing the change of variables $\tilde{\varphi}_{\mathbf{q}}
=
\arcsin[2 (\tilde{\kappa_z} \tilde{k}_z 
+
\tilde{\mathbf{Q}}_{\mathbf{q},\text{R}}
\cdot
\tilde{\mathbf{Q}}_{\mathbf{q},\text{I}}) / \tilde{\varkappa}_{\mathbf{q}}]$, 
$\tilde{\vartheta}_{\mathbf{q}}
=
\arcsin(2 k_z \tilde{Q}_{\mathbf{q},\text{R}}/\nu_{\mathbf{q}})$ and
$\tilde{\vartheta}_{\mathbf{q}}^{\prime} 
= 
\arcsin(2 k_z \tilde{Q}_{\mathbf{q},\text{I}}/\nu_{\mathbf{q}}^{\prime})$ with $\tilde{\mathbf{Q}}_{\mathbf{q}, \text{R}}
=
\text{Re}(\tilde{\mathbf{Q}}_{\mathbf{q}})$, 
$\tilde{\mathbf{Q}}_{\mathbf{q}, \text{I}}
=
\text{Im}(\tilde{\mathbf{Q}}_{\mathbf{q}})$ and
\begin{subequations}
\begin{align}
\tilde{\varkappa}_{\mathbf{q}}^2
&=
(\tilde{Q}_{\mathbf{q},\text{R}}^2 
-
\tilde{Q}_{\mathbf{q},\text{I}}^2
-
\tilde{\kappa}_z^2 + \tilde{k}_z^2)^2
+
4(\tilde{\mathbf{Q}}_{\mathbf{q},\text{R}} 
\cdot
\tilde{\mathbf{Q}}_{\mathbf{q},\text{I}}
+
\tilde{\kappa}_z \tilde{k}_z)^2,
\\
\nu_{\mathbf{q}}^2 
&=
[\tilde{Q}_{\mathbf{q},\text{R}}^2 
-
\tilde{Q}_{\mathbf{q},\text{I}}^2
-
\tilde{\kappa}_z^2 - \tilde{k}_z^2
-
2 \tilde{k}_z \text{Im}(\tilde{\xi}_{0})]^2
+
(2 k_z \tilde{Q}_{\mathbf{q},\text{R}})^2,
\\
\nu_{\mathbf{q}}^{\prime 2} 
&=
4[ \tilde{\mathbf{Q}}_{\mathbf{q},\text{R}} 
\cdot
\tilde{\mathbf{Q}}_{\mathbf{q},\text{I}}
-
\tilde{\kappa}_z \text{Im}(\tilde{\xi}_{0}
)]^2
+
(2 k_z \tilde{Q}_{\mathbf{q},\text{I}})^2,
\end{align}
\end{subequations}
the dressed reflection matrix is reexpressed as
\begin{equation}
\label{Seq:R_tilde}
\hat{\tilde{R}}_{\mathbf{q}}
=
\frac{e^{i \tilde{\varphi}_{\mathbf{q}}}}{\tilde{\varkappa}_{\mathbf{q}}}
\left(
\nu_{\mathbf{q}}
e^{i \tilde{\vartheta} _{\mathbf{q}}
\hat{\boldsymbol{\sigma}}\cdot
\tilde{\mathbf{n}}_{\mathbf{q},\text{R}}}
+
i \nu^{\prime}_{\mathbf{q}}
e^{i \tilde{\vartheta}^{\prime} _{\mathbf{q}}
\hat{\boldsymbol{\sigma}}\cdot 
\tilde{\mathbf{n}}_{\mathbf{q},\text{I}}}
\right),
\end{equation}
which is Eq.~(\ref{R_tilde}) in the main text. Here, we have introduced the unit vectors $\tilde{\mathbf{n}}_{\mathbf{q},\text{R}}
=
\tilde{\mathbf{Q}}_{\mathbf{q},\text{R}}/
\tilde{Q}_{\mathbf{q},\text{R}}$ and 
$\tilde{\mathbf{n}}_{\mathbf{q},\text{I}}
=
\tilde{\mathbf{Q}}_{\mathbf{q},\text{I}}/
\tilde{Q}_{\mathbf{q},\text{I}}$, corresponding to the real and imaginary parts of the vector $\tilde{\mathbf{Q}}_{\mathbf{q}}$, respectively. As a useful consistency check, it is worth noting that in the limit of vanishing disorder, $\nu_{\mathbf{q}}, \tilde{\varkappa}_{\mathbf{q}} \rightarrow \varkappa_{\mathbf{q}}$, while $\nu_{\mathbf{q}}^{\prime} \rightarrow 0$, hence $\hat{\tilde{R}}_{\mathbf{q}} \rightarrow \hat{R}_{\mathbf{q}}$ and Eq.~(\ref{Seq:R_hat2}) is reobtained.

Before moving on, we present a brief derivation of the disorder averaged irreducible self-energy, which  is useful when doing explicit calculations of physical quantities. Inserting Eq.~(\ref{gR_hat_nm}) into Eq.~(\ref{Sigma_tilde}) and using the previously introduced change of variables, we find
\begin{equation}
\hat{\tilde{\xi}} \left(\epsilon\right)
=
\frac{1}{8 \pi^2 i} \eta_{\gamma} \int d^2\mathbf{q}\, \frac{1}{k_z} \left(1 + \hat{R}_{\mathbf{q}} \right),
\end{equation}
with $\eta_{\gamma}$ the dimensionless strength of the interfacial impurity interaction. Keeping terms up to first order in the Rashba SOC and exchange constants in the reflection matrix $\hat{R}_{\mathbf{q}}$, given by Eq.~(\ref{Seq:R_hat1}), and integrating up to the cutoff $q_{\text{max}}=k$, we obtain the result
\begin{subequations}
\label{xi_mu}
\begin{gather}
\tilde{\xi}_0
\simeq
-\frac{1}{6} \eta_{\gamma} k
\left[ \frac{1- \left(1 - \Delta^2\right)^{\frac{3}{2}}}{\Delta} + i \Delta^2 \right],
\\
\tilde{\bs{\xi}}
\simeq
-\frac{1}{8} \eta_{\gamma} \eta_{ex} k
\left[ 2 \Delta^2 \left( 1- \Delta^2 \right)
+ i \left( \sin^{-1}\left(\Delta\right) - \Delta \left(1- 2 \Delta^2 \right) \sqrt{1- \Delta^2} \right)\right] \mathbf{m},
\end{gather}
\end{subequations}
where $\eta_{ex} \equiv \xi_{ex}/k$ is the dimensionless exchange constant and $\Delta= k/k_b= \sqrt{|\epsilon|/V_b}$ is a dimensionless quantity that measures the ratio of the electron's wavelength $k=\sqrt{2m |\epsilon| /\hbar^2}$ to the wavelength associated with the potential barrier, $k_b=\sqrt{2m V_b/\hbar^2}$. Note that at the Fermi level, we have $\eta_{ex} = \xi_{ex}/k_F$ and $\Delta=k_F/k_b$.

\section{Quadratic Response Theory}
\label{appendixB}

As magnetotransport effects that respond quadratically to an applied electric field, unidirectional magnetoresistances (UMRs) and nonlinear Hall effects cannot be captured within the formalism of linear response theory. In the language of diagrammatic response theory, this implies that, instead of the well known two-photon bubble diagrams that are widely used to calculate linear responses \cite{mahan2000many, bruus2004many}, a formal quantum calculation of UMRs and nonlinear Hall effects mandates the use of response diagrams that have \textit{three} external photon legs, namely triangle diagrams, three-photon bubble diagrams and three-photon-vertex diagrams \cite{parker2019diagrammatic,du2021quantum, rostami2021gauge}.

In this section, we present the quadratic conductivity tensors--both with and without interfacial disorder--in mixed real and momentum space, which is required for the study of quantum interference of electrons at the interface of a bilayer system. Then, through a physically relevant approximation, we derive an analytical expression for the conductivity tensor and show that it may be expressed entirely in terms of the interference velocity--even in the presence of bulk and interfacial disorder. This confirms the role of quantum interference in generating the longitudinal and transverse QUMRs and suggests the robustness of these nonlinear magnetoresistances against disorder effects.

\subsection{Conductivities without Interfacial Disorder}

\begin{figure}[tph]
    \includegraphics[width=.6\linewidth]{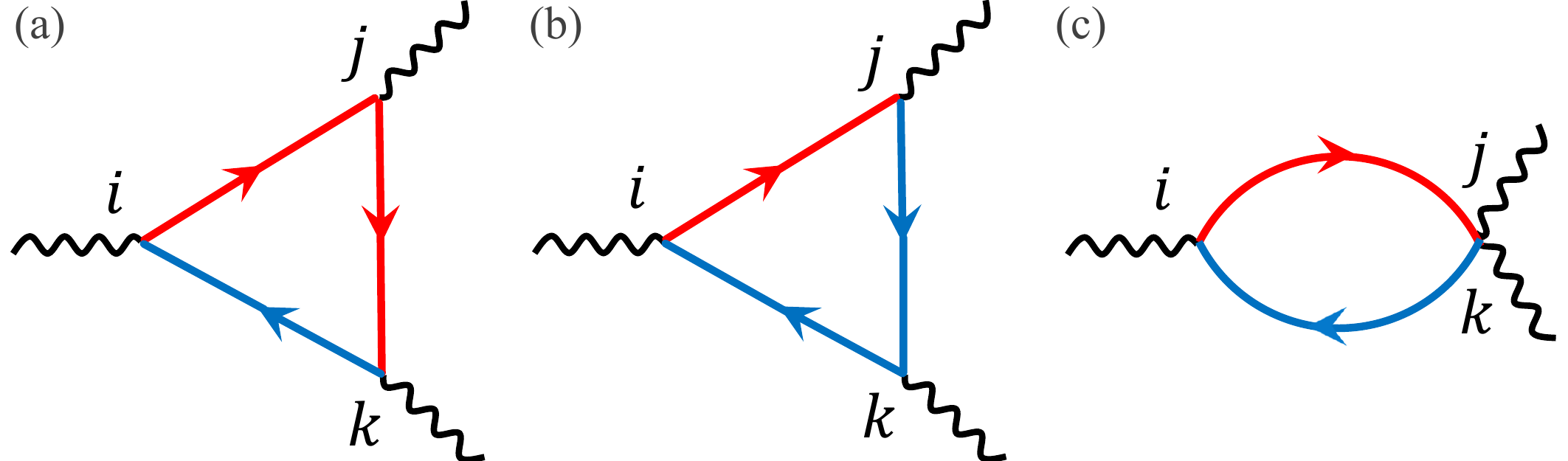}
    \caption{Diagrammatic structure of the (bare) quadratic responses in the absence of interfacial disorder. Together, figures (a)-(c), along with their $j \leftrightarrow k$ counterparts, comprise the undressed conductivity tensor $\sigma_{ijk}$. Here, red (blue) arrowed lines represent retarded (advanced) propagators, while black wavy lines are external photon legs.}
    \label{figS2}%
\end{figure}

Let us first consider the simpler case where interfacial disorder is absent. In general, both the Fermi sea and the Fermi surface will contribute to the nonlinear transport. However, in the weak disorder limit, we may neglect the Fermi sea contribution and focus on the Fermi surface contribution \cite{mahan2000many}. For a Hamiltonian such as that given by Eqs.~(\ref{Seq:H_hat}), which is at most quadratic in momentum, as shown in Fig.~\ref{figS2}, the local quadratic conductivity thus consists of terms arising from triangle diagrams, $\sigma_{ijk}^{(a)} (\mathbf{r})$, as well as from three-photon bubble diagrams, $ \sigma_{ijk}^{(b)} (\mathbf{r})$, so that the total conductivity is
\begin{equation}
\sigma_{ijk} (\mathbf{r})
=
\sigma_{ijk}^{(a)} (\mathbf{r}) + \sigma_{ijk}^{(b)} (\mathbf{r}),
\end{equation}

\noindent where the real-space representations of the individual conductivities read \cite{du2021quantum}
\begin{subequations}
\label{Seq:sigma_ijk_r}
\begin{align}
\begin{split}
\sigma_{ijk}^{(a)}(\mathbf{r})
=&
-\frac{e^3 \hbar^2}{\pi} \int_{-\infty}^{\infty} d\epsilon \,\pd_{\epsilon} f
\int d\mathbf{r}_1 \cdots \int d\mathbf{r}_5
\\
&\hspace{0.09\linewidth}\times \text{Im} \left\{\text{Tr} \left[
\braket{\mathbf{r}|\hat{v}_i |\mathbf{r}_1}
\pd_{\epsilon} \hat{G}^R\left(\mathbf{r}_1, \mathbf{r}_2; \epsilon \right) 
\braket{\mathbf{r}_2| \hat{v}_j |\mathbf{r}_3}
\hat{G}^R\left(\mathbf{r}_3, \mathbf{r}_4; \epsilon \right)
\braket{\mathbf{r}_4| \hat{v}_k |\mathbf{r}_5}
\hat{G}^A\left(\mathbf{r}_5, \mathbf{r}; \epsilon \right)
\right]\right\}
+
\left(j \leftrightarrow k \right),
\end{split}
\\
\sigma_{ijk}^{(b)}(\mathbf{r})
=
&-\frac{e^3 \hbar^2}{2 \pi} 
\int_{-\infty}^{\infty} d\epsilon \,\pd_{\epsilon} f
\int d\mathbf{r}_1 \cdots \int d\mathbf{r}_3
\,\text{Im} \left\{\text{Tr} 
\left[
\braket{\mathbf{r}|\hat{v}_i |\mathbf{r}_1}
\pd_{\epsilon}
\hat{G}^R\left(\mathbf{r}_1, \mathbf{r}_2; \epsilon \right)
\braket{\mathbf{r}_2|\hat{v}_{jk} |\mathbf{r}_3}
\hat{G}^A\left(\mathbf{r}_3, \mathbf{r}; \epsilon \right)
\right]\right\}
+
\left(j \leftrightarrow k \right),
\end{align}
\end{subequations}
with $\pd_{\epsilon} \equiv \pd/\pd \epsilon$ and the advanced propagator related to the retarded one as
\begin{equation}
\hat{G}^A \left(\mathbf{r}^{\prime}, \mathbf{r}; \epsilon \right)
=
\left[\hat{G}^R \left(\mathbf{r}, \mathbf{r}^{\prime}; \epsilon \right)\right]^{\dagger}.
\end{equation}
In Eqs.~(\ref{Seq:sigma_ijk_r}), the replacement $j \leftrightarrow k$ ensures the intrinsic permutation symmetry of the quadratic response function. And the first- and second-order velocity operators are given by
\begin{subequations}
\begin{align}
\hat{v}^i
&=
\frac{1}{\hbar} \partial_{\mathbf{k}}^i \hat{H}_0,
\\
\hat{v}^{ij}
&=
\frac{1}{\hbar^2} \partial_{\mathbf{k}}^i \partial_{\mathbf{k}}^j \hat{H}_0,
\end{align}
\end{subequations}
with $\partial_{\mathbf{k}}^i \equiv \partial/\partial k_i$. For in-plane momenta, this leads to the real-space representations
\begin{subequations}
\begin{align}
\braket{\mathbf{r}|\hat{v}^i |\mathbf{r}_1}
&=
\frac{\hbar}{m}
\sum_{\mathbf{q}} q^i e^{i \mathbf{q} \cdot \left(\bs{\rho} - \bs{\rho}_1 \right)} \delta \left(z-z_1 \right),
\\
\braket{\mathbf{r}|\hat{v}^{ij} |\mathbf{r}_1}
&=
\frac{1}{m}
\delta^{ij} \delta \left(\mathbf{r} - \mathbf{r}_1 \right).
\end{align}
\end{subequations}
Applying Eq.~(\ref{gR_hat_z}), the conductivities now read
\begin{subequations}
\label{sigma_ijk}
\begin{align}
\begin{split}
\sigma_{ijk}^{(a)}(z)
=&
-\frac{2 e^3 \hbar^5}{\pi m^{3}}
\int \frac{d^2\mathbf{q}}{\left( 2\pi \right)^2} \int_{-\infty}^{\infty} d\epsilon \,\pd_{\epsilon} f
\int_0^{\infty} dz^{\prime} \int_0^{\infty} dz^{\dprime} \,q_i q_j q_k
\text{Im} \left\{\text{Tr} \left[
\pd_{\epsilon} \hat{g}^R_{\mathbf{q}}\left(z, z^{\prime}; \epsilon \right)
\hat{g}^R_{\mathbf{q}}\left(z^{\prime}, z^{\dprime}; \epsilon \right)
\hat{g}^A_{\mathbf{q}}\left(z^{\dprime}, z ;\epsilon \right)\right]\right\},
\end{split}
\\
\sigma_{ijk}^{(b)}(z)
=
&-\frac{e^3 \hbar^3}{\pi m^2} \int \frac{d^2\mathbf{q}}{\left( 2\pi \right)^2} \int_{-\infty}^{\infty} d\epsilon \,\pd_{\epsilon} f
\int_0^{\infty} dz^{\prime} \,q_i \delta_{jk}
\,\text{Im} \left\{\text{Tr} \left[
\pd_\epsilon \hat{g}^R_{\mathbf{q}} \left(z, z^{\prime}; \epsilon \right)
\hat{g}^A_{\mathbf{q}} \left(z^{\prime}, z; \epsilon \right)\right]\right\}.
\end{align}
\end{subequations}
Inserting Eq.~(\ref{gR_hat_nm}) into Eqs.~(\ref{sigma_ijk}), the analytical form of the nonlinear conductivities in terms of the reflection matrix may be obtained, which turns out to be rather cumbersome. In the limit $l_n\gg \lambda_F$, with $\lambda_F=2\pi/k_F$ the Fermi wavelength in the NM layer, the highest-order terms in the mean free path will dominate the transport and the conductivities may be approximated as
\begin{subequations}
\label{Seq:sigma_ijk_ab_ballistic}
\begin{align}
\sigma_{ijk}^{(a)}(z)
=
&-\frac{2 e^3 m}{\pi \hbar^3} l_n^2 \int \frac{d^2\mathbf{q}}{\left( 2\pi \right)^2} \int_{-\infty}^{\infty} d\epsilon \,\pd_{\epsilon} f \,q_i q_j q_k
\left(
\frac{2k^2 + k_z^2}{k^4 k_z^3}
-
\frac{\hbar^2}{m}
\frac{1}{k^2 k_z} \pd_{\epsilon}
\right)
\text{Re} \left(e^{2i k_z z} \text{Tr} \hat{R}_{\mathbf{q}}\right),
\\
\sigma_{ijk}^{(b)}(z)
=
&-\frac{2 e^3 m}{\pi \hbar^3} l_n^2 
\int \frac{d^2\mathbf{q}}{\left( 2\pi \right)^2} 
\int_{-\infty}^{\infty} d\epsilon \,\pd_{\epsilon} f \,
q_i \delta_{jk} \frac{1}{k^2 k_z}
\text{Re} \left(e^{2i k_z z} \text{Tr} \hat{R}_{\mathbf{q}}\right).
\end{align}
\end{subequations}
Using the approximation $\pd_{\epsilon}f \simeq - \delta(\epsilon- \epsilon_F)$ and the expression for the interference velocity at the Fermi level
\begin{equation}
\Braket{\hat{\bs{v}} (\mathbf{q},z)}_{I-R}
=
\frac{2 \hbar \mathbf{q}}{m}
\text{Re} \left(e^{2i k_{z,F} z} \text{Tr} \hat{R}_{\mathbf{q}}\right),
\end{equation}
with $k_{z,F}=\sqrt{k_{F}^{2}-q^{2}}$, Eqs.~(\ref{Seq:sigma_ijk_ab_ballistic}) are recast in the form
\begin{equation}
\label{Seq:sigma_ijk_ballistic}
\sigma_{ijk}\left( z ; \mathbf{m} \right)
=
\frac{2 e^3 m^{2}}{\pi \hbar^4 k_F^2} l_n^2
\int \frac{d^2\mathbf{q}}{\left( 2\pi \right)^2}
\frac{1}{k_{z,F}} 
\left[
q_j q_k
\left(\frac{2}{k_{z,F}^2} + \frac{1}{k_F^2} - \frac{\hbar^2}{m}\pd_{\epsilon_F}\right)
+
\delta_{jk}
\right]
\Braket{\hat{v}_{i}(\mathbf{q},z)}_{I-R}.
\end{equation}
In this form, we readily see the central role played by the interference velocity $\braket{\hat{\bs{v}}({\mathbf{q}},z)}_{I-R}$ in generating the nonlinear response. Next, we study the effect of the interfacial disorder on the nonlinear response.  

\subsection{Including Interfacial Disorder}

\begin{figure}[tph]
    \includegraphics[width=.9\linewidth]{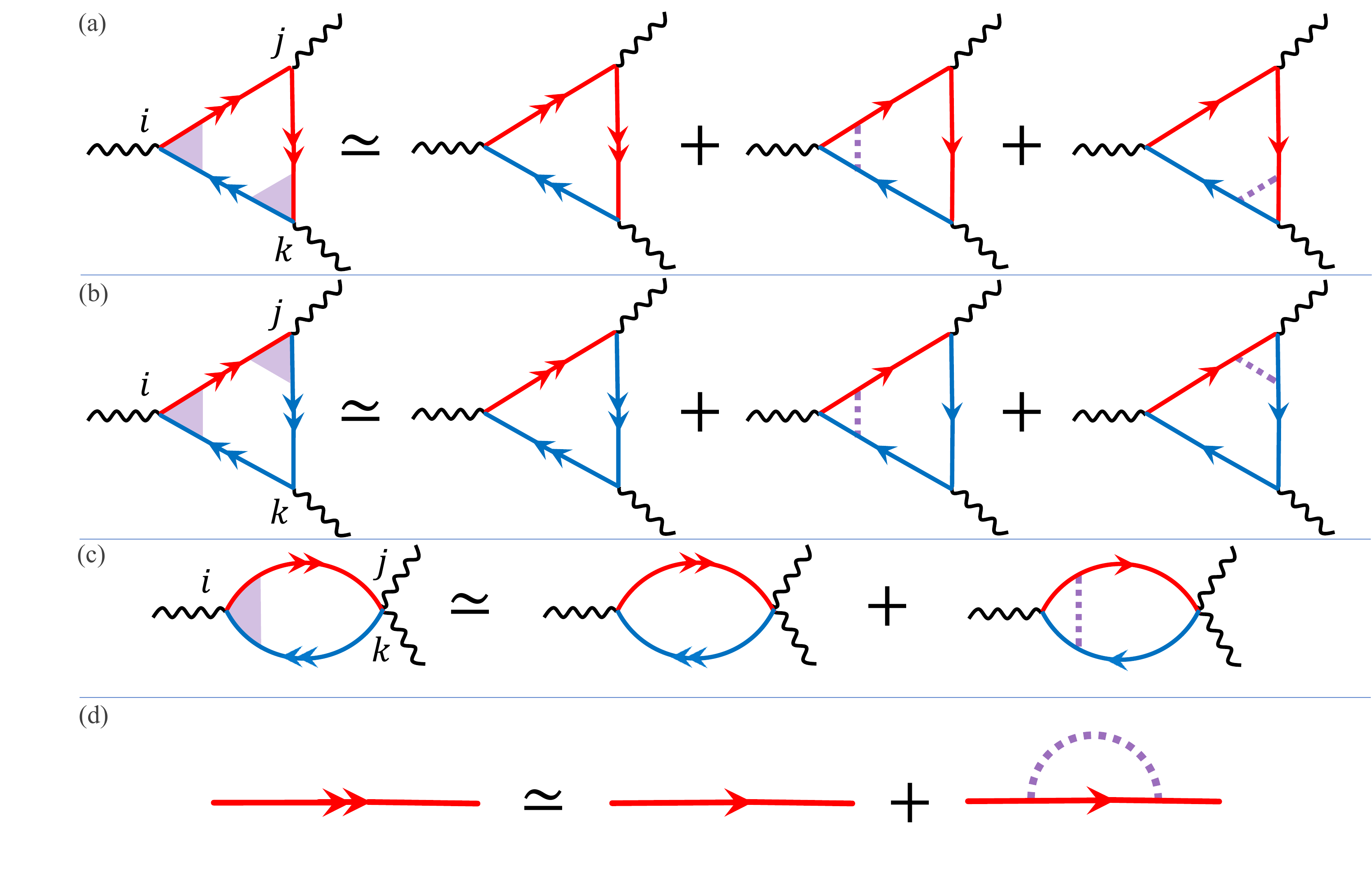}
    \caption{(a-c) Leading-order contributions to the dressed conductivity tensor, where double-arrowed lines represent dressed propagators. The purple shaded areas correspond to dressed vertices, while the leading-order disorder vertex corrections are represented by dotted purple lines. (d) The dressed propagator within the Born approximation.}
    \label{figS3}%
\end{figure}

We now consider the contributions of impurities at the bilayer interface. In the presence of interfacial disorder, the propagator is no longer diagonal in momentum space and the conductivities generalize to
\begin{subequations}
\label{sigma_ijk_disorder}
\begin{align}
\begin{split}
\tilde{\sigma}_{ijk}^{(a)}(z)
&=
-\frac{e^3 \hbar^5}{\pi m^{3}} 
\left[\int \frac{d^2\mathbf{q}}{\left( 2\pi \right)^2}
\cdots
\int \frac{d^2\mathbf{q}^{\tprime}}{\left( 2\pi \right)^2}
\right]
\int_{-\infty}^{\infty} d\epsilon \,\pd_{\epsilon} f
\int_0^{\infty} dz^{\prime} \int_0^{\infty} dz^{\dprime} \,q_i q^{\prime}_j q^{\dprime}_k
\\
&\hspace{0.34\linewidth}\times \text{Im} \left\{\text{Tr} 
\left[
\pd_{\epsilon} 
\hat{\tilde{g}}^R_{\mathbf{q} \mathbf{q}^{\prime}}\left(z, z^{\prime}; \epsilon \right) 
\hat{\tilde{g}}^R_{\mathbf{q}^{\prime} \mathbf{q}^{\dprime}} \left(z^{\prime}, z^{\dprime}; \epsilon \right) 
\hat{\tilde{g}}^A_{ \mathbf{q}^{\dprime} \mathbf{q}^{\tprime}} \left(z^{\dprime}, z; \epsilon \right)\right]\right\}
+
\left( j \leftrightarrow k\right),
\end{split}
\\
\begin{split}
\tilde{\sigma}_{ijk}^{(b)}(z)
&=
-\frac{e^3 \hbar^3}{2 \pi m^2} 
\left[\int \frac{d^2\mathbf{q}}{\left( 2\pi \right)^2}
\cdots
\int \frac{d^2\mathbf{q}^{\dprime}}{\left( 2\pi \right)^2}
\right]
\int_{-\infty}^{\infty} d\epsilon \,\pd_{\epsilon} f
\int_0^{\infty} dz^{\prime} \,q_i \delta_{jk}
\text{Im} \left\{\text{Tr} 
\left[
\pd_{\epsilon} 
\hat{\tilde{g}}^R_{\mathbf{q} \mathbf{q}^{\prime}} \left(z, z^{\prime}; \epsilon \right) 
\hat{\tilde{g}}^A_{\mathbf{q}^{\prime} \mathbf{q}^{\dprime}} \left(z^{\prime}, z; \epsilon \right) \right]\right\}
+
\left( j \leftrightarrow k\right).
\end{split}
\end{align}
\end{subequations}
In order to calculate the specular and diffuse contributions of Eqs.~(\ref{sigma_ijk_disorder}), one needs to calculate configurational averages of products of two and three propagators, which we symbolically express in the condensed notation $\braket{\hat{\tilde{G}}^2}$ and $\braket{\hat{\tilde{G}}^3}$. Let us also reexpress Eq.~(\ref{GR_imp}) in the condensed form
\begin{equation}
\hat{\tilde{G}}
=
\hat{G} + \hat{G} \hat{V}^{\text{imp}} \hat{\tilde{G}}.
\end{equation}
We then have
\begin{subequations}
\label{condensed}
\begin{align}
\Braket{\hat{\tilde{G}}^2}
&=
\Braket{\left(\hat{G} + \hat{G} \hat{V}^{\text{imp}} \hat{\tilde{G}}\right)^2}
\simeq
\Braket{\hat{\tilde{G}}}^2
+
\Braket{\left(\hat{G} \hat{V}^{\text{imp}} \hat{G} \right)^2},
\\
\Braket{\hat{\tilde{G}}^3}
&=
\Braket{\left(\hat{G} + \hat{G} \hat{V}^{\text{imp}} \hat{\tilde{G}}\right)^3}
\simeq
\Braket{\hat{\tilde{G}}}^3
+
\Braket{\left(\hat{G} \hat{V}^{\text{imp}} \hat{G}\right)^2} \hat{G}
+
\Braket{\left(\hat{G} \hat{V}^{\text{imp}} \hat{G} \right) \hat{G} \left(\hat{G} \hat{V}^{\text{imp}} \hat{G} \right)}
+
\hat{G} \Braket{\left(\hat{G} \hat{V}^{\text{imp}} \hat{G} \right)^2},
\end{align}
\end{subequations}
where we have retained only the leading order contributions to the vertex corrections. In Eqs.~(\ref{condensed}), terms containing separately averaged propagators, $\braket{\hat{\tilde{G}}}^2$ and $\braket{\hat{\tilde{G}}}^3$, correspond to momentum-preserving scatterings. Hence, they constitute the specular contribution to the conductivity tensor. The vertex corrections, on the other hand, contain momentum-mixing terms and contribute to the diffuse scattering. A summary of these contributions are presented in Fig.~\ref{figS3}, in which we diagrammatically highlight the approximations to the renormalized propagators and velocity vertex functions. 

As we neglect mixing terms--which appear at higher order in the interfacial disorder parameter--the dressed conductivity tensor $\tilde{\sigma}_{ijk}$ may be expressed as 
\begin{equation}
\tilde{\sigma}_{ijk}
=
\tilde{\sigma}_{ijk}^{\text{spec}}
+
\tilde{\sigma}_{ijk}^{\text{diff}},
\end{equation}
where, here and henceforth, we omit the ensemble-averaging brackets $\braket{\cdots}$ for simplicity.
Below, we present explicit calculations of the specular and diffuse contributions to the nonlinear response tensor.

\subsubsection{Specular Contribution}

In the presence of specular scattering alone, as shown in Eqs.~(\ref{condensed}), the dressed conductivity is obtained in a straightforward manner by simply replacing the bare propagators in Eqs.~(\ref{sigma_ijk}) with their dressed counterparts as
\begin{subequations}
\label{Seq:sigma_ijk_spec}
\begin{align}
\begin{split}
\tilde{\sigma}_{ijk}^{(a),\text{spec}}(z)
&=
-\frac{2 e^3 \hbar^5}{\pi m^{3}} \int \frac{d^2\mathbf{q}}{\left( 2\pi \right)^2} \int_{-\infty}^{\infty} d\epsilon \,\pd_{\epsilon} f
\int_0^{\infty} dz^{\prime} \int_0^{\infty} dz^{\dprime} \,q_i q_j q_k
\text{Im} \left\{\text{Tr} \left[
\pd_{\epsilon} \hat{\tilde{g}}^R_{\mathbf{q}} \left(z, z^{\dprime}; \epsilon \right)
\hat{\tilde{g}}^R_{\mathbf{q}} \left(z^{\dprime}, z^{\prime}; \epsilon \right) 
\hat{\tilde{g}}^A_{\mathbf{q}} \left(z^{\prime}, z; \epsilon \right)\right]\right\},
\end{split}
\\
\tilde{\sigma}_{ijk}^{(b),\text{spec}}(z)
&=
-\frac{e^3 \hbar^3}{\pi m^2}
\int \frac{d^2\mathbf{q}}{\left( 2\pi \right)^2}
\int_{-\infty}^{\infty} d\epsilon \,\pd_{\epsilon} f
\int_0^{\infty} dz^{\prime} \,q_i \delta_{jk}
\,\text{Im} \left\{\text{Tr} \left[
\pd_\epsilon \hat{\tilde{g}}^R_{\mathbf{q}} \left(z, z^{\prime}; \epsilon \right)
\hat{\tilde{g}}^A_{\mathbf{q}} \left(z^{\prime}, z; \epsilon \right)\right]\right\}.
\end{align}
\end{subequations}
Note that with the approximation $\pd_{\epsilon}f \simeq - \delta(\epsilon- \epsilon_F)$, Eqs.~(\ref{Seq:sigma_ijk_spec}) reduce to Eqs.~(\ref{eq:sigma_ijk}) presented in the main text.

Similarly, the analytical approximation is obtained by the simple replacement of the reflection matrix $\hat{R}_{\mathbf{q}}$ in Eqs.~(\ref{Seq:sigma_ijk_ab_ballistic}) with its disordered counterpart $\hat{\tilde{R}}_{\mathbf{q}}$, given by Eq.~(\ref{R_hat_tilde})--or equivalently, Eq.~(\ref{Seq:R_tilde}). We thus arrive at the generalized result given by Eq.~(\ref{sigma_ijk_ballistic})
\begin{equation}
\label{sigma_ijk_ballistic_SM}
\tilde{\sigma}_{ijk}^{\text{spec}} \left( z ; \mathbf{m} \right)
=
\frac{2 e^3 m^{2}}{\pi \hbar^4 k_F^2} l_n^2
\int \frac{d^2\mathbf{q}}{\left( 2\pi \right)^2}
\frac{1}{k_{z,F}} 
\left[
q_j q_k
\left(\frac{2}{k_{z,F}^2} + \frac{1}{k_F^2} - \frac{\hbar^2}{m}\pd_{\epsilon_F}\right)
+
\delta_{jk}
\right]
\Braket{\hat{\tilde{v}}_{i}(\mathbf{q},z)}_{I-R},
\end{equation}
where the dressed interference velocity is given by
\begin{equation}
\Braket{\hat{\tilde{\bs{v}}} (\mathbf{q},z)}_{I-R}
=
\frac{2 \hbar \mathbf{q}}{m}
\text{Re} \left(e^{2i k_{z,F} z} \text{Tr} \hat{\tilde{R}}_{\mathbf{q}}\right).
\end{equation}

\subsubsection{Diffuse Correction}

We now present the diffuse corrections to the conductivity tensor, which are all the diagrams in Fig.~\ref{figS3} (along with their $j \leftrightarrow k$ counterparts) that include vertex corrections. Following Eqs.~(\ref{condensed}), we conclude that the diffuse corrections corresponding to dressed triangle diagrams, $\tilde{\sigma}_{ijk}^{(a),\text{diff}}$, and dressed three-photon bubble diagrams,  $\tilde{\sigma}_{ijk}^{(b),\text{diff}}$, can be expressed as
\begin{subequations}
\label{Seq:sigma_ijk_diff}
\begin{align}
\begin{split}
\tilde{\sigma}_{ijk}^{(a),\text{diff}}(z)
=&
-\frac{e^3 \hbar^5}{\pi m^{3}}
\int_{-\infty}^{\infty} d\epsilon \,\pd_{\epsilon} f\,
\text{Im}
\left\{ \text{Tr}
\left[ \hat{\mathcal{P}}_{ijk}\left( z;\epsilon\right) \right] \right\}
+
\left( j \leftrightarrow k \right),
\end{split}
\\
\tilde{\sigma}_{ijk}^{(b),\text{diff}}(z)
=
&-\frac{e^3 \hbar^3}{2 \pi m^2} 
\int_{-\infty}^{\infty} d\epsilon \,\pd_{\epsilon} f\,
\text{Im}
\left\{ \text{Tr}
\left[ \hat{\mathcal{S}}_{ijk}\left( z;\epsilon\right) \right] \right\}
+
\left( j \leftrightarrow k \right),
\end{align}
\end{subequations}
where
\begin{equation}
\hat{\mathcal{P}}_{ijk}\left( z;\epsilon\right)
=
\hat{\mathcal{P}}_{ijk}^{(1)}\left( z;\epsilon\right)
+
\hat{\mathcal{P}}_{ijk}^{(2)}\left( z;\epsilon\right)
+
\hat{\mathcal{P}}_{ijk}^{(3)}\left( z;\epsilon\right),
\end{equation}
with
\begin{subequations}
\label{F_tilde_I}
\begin{align}
\begin{split}
\hat{\mathcal{P}}_{ijk}^{(1)}\left( z;\epsilon\right)
=
&\left(\frac{\hbar^2}{2m}\right)^2 \eta_{\gamma}
\int \frac{d^2\mathbf{q}}{\left( 2\pi \right)^2}
\int \frac{d^2\mathbf{q}^{\prime}}{\left( 2\pi \right)^2}
\int_0^{\infty} dz^{\prime}
\int_0^{\infty} dz^{\dprime}\, q^{\prime}_i q_j q_k
\\
&\hspace{0.22\linewidth}\times 
\pd_{\epsilon}\left[
\hat{g}^R_{\mathbf{q}^{\prime}} \left(z,0;\epsilon\right) 
\hat{g}^R \left(0,z^{\prime},\mathbf{q};\epsilon\right)
\right]
\hat{g}^R_{\mathbf{q}} \left(z^{\prime},z^{\dprime};\epsilon\right)
\hat{g}^A_{\mathbf{q}} \left(z^{\dprime},0;\epsilon \right)
\hat{g}^A_{\mathbf{q}^{\prime}} \left(0,z;\epsilon\right),
\end{split}
\\
\begin{split}
\hat{\mathcal{P}}_{ijk}^{(2)}\left( z;\epsilon\right)
=
&\left(\frac{\hbar^2}{2m}\right)^2 \eta_{\gamma}
\int \frac{d^2\mathbf{q}}{\left( 2\pi \right)^2}
\int \frac{d^2\mathbf{q}^{\prime}}{\left( 2\pi \right)^2}
\int_0^{\infty} dz^{\prime}
\int_0^{\infty} dz^{\dprime}\, q_i q^{\prime}_j q_k
\\
&\hspace{0.22\linewidth}\times 
\pd_{\epsilon}\left[
\hat{g}^R_{\mathbf{q}} \left(z,0;\epsilon\right) 
\hat{g}^R_{\mathbf{q}^{\prime}} \left(0,z^{\prime};\epsilon\right)
\right]
\hat{g}^R_{\mathbf{q}^{\prime}}\left(z^{\prime},0;\epsilon\right)
\hat{g}^R_{\mathbf{q}} \left(0,z^{\dprime};\epsilon\right)
\hat{g}^A_{\mathbf{q}} \left(z^{\dprime},z;\epsilon\right),
\end{split}
\\
\begin{split}
\hat{\mathcal{P}}_{ijk}^{(3)}\left( z;\epsilon\right)
=
&\left(\frac{\hbar^2}{2m}\right)^2 \eta_{\gamma}
\int \frac{d^2\mathbf{q}}{\left( 2\pi \right)^2}
\int \frac{d^2\mathbf{q}^{\prime}}{\left( 2\pi \right)^2}
\int_0^{\infty} dz^{\prime}
\int_0^{\infty} dz^{\dprime}\, q_i q_j q^{\prime}_k
\\
&\hspace{0.22\linewidth}\times 
\pd_{\epsilon}\left[
\hat{g}^R_{\mathbf{q}} \left(z,z^{\prime};\epsilon\right) 
\right]
\hat{g}^R_{\mathbf{q}} \left(z^{\prime},0;\epsilon\right)
\hat{g}^R_{\mathbf{q}^{\prime}} \left(0, z^{\dprime};\epsilon\right)
\hat{g}^A_{\mathbf{q}^{\prime}} \left(z^{\dprime},0;\epsilon \right)
\hat{g}^A_{\mathbf{q}} \left(0,z;\epsilon\right),
\end{split}
\end{align}
\end{subequations}
and
\begin{equation}
\label{F_tilde_II}
\begin{split}
\hat{\mathcal{S}}_{ijk}\left( z;\epsilon\right)
=
&\left(\frac{\hbar^2}{2m}\right)^2 \eta_{\gamma}
\int \frac{d^2\mathbf{q}}{\left( 2\pi \right)^2}
\int \frac{d^2\mathbf{q}^{\prime}}{\left( 2\pi \right)^2}
\int_0^{\infty} dz^{\prime}\, q_i \delta_{jk}
\pd_{\epsilon}\left[
\hat{g}^R_{\mathbf{q}} \left(z,0;\epsilon\right) 
\hat{g}^R_{\mathbf{q}^{\prime}} \left(0,z^{\prime};\epsilon\right)
\right]
\hat{g}^A_{\mathbf{q}^{\prime}} \left(z^{\prime},0;\epsilon\right)
\hat{g}^A_{\mathbf{q}} \left(0,z;\epsilon\right).
\end{split}
\end{equation}
While an analytical approximation to Eqs.~(\ref{Seq:sigma_ijk_diff}) is rather intractable, a numerical calculation reveals that the diffuse correction to the UMR coefficients tends to counteract the specular corrections arising from the interfacial disorder. However, the diffuse contributions turn out to be at least two orders of magnitude smaller than the overall UMR coefficient strengths. Thus, they will not affect the main physical results and may safely be neglected in the present study.

\twocolumngrid
\bibliographystyle{my-aps-style}
\bibliography{qumr}
\end{document}